# Triaxial nuclei and analytical solutions of the conformable fractional Bohr Hamiltonian with some exponential-type potentials

M. M. Hammad[1,2], M. M. Yahia[2], Dennis Bonatsos[1]

[1]Institute of Nuclear and Particle Physics, National Centre for Scientific Research "Demokritos", GR- 15310 Aghia Paraskevi, Attiki, Greece
[2]Faculty of Science, Mathematics Department, Damanhour University, Egypt

**Abstract**

New approximate analytical solutions have been obtained for the conformable fractional collective Bohr Hamiltonian suitable for triaxial nuclei, with the harmonic oscillator in $\gamma$-part of the collective potential and different exponential-type potentials (namely Morse, Tietz-Hua, and multi-parameter exponential-type potential) in $\beta$-part. The conformable fractional Nikiforov-Uvarov approach is used to derive closed analytical formulas for energy spectra and wave functions. The relation between the conformable fractional spectra of the three potentials and the Z(5) spectrum has been studied. The evolution of the energy spectra as a function of the potential parameters has been investigated for the three potentials. The normalized $B(E2)$ transition rates and spectra have been calculated and compared with the experimental data and theoretical predictions of Kratzer potential. The predictions can describe well the experimental results for $^{114,116}$Pd, $^{126,128}$Xe, and $^{192,194}$Pt isotopes.

**Keywords:** Bohr Hamiltonian, triaxial nuclei, quantum shape phase transitions, Z(5) model, Morse potential, Tietz-Hua potential, multi-parameter exponential-type potential, conformable fractional calculus, Nikiforov-Uvarov method.

## 1. Introduction

In the last decade, nuclear quantum shape phase transitions (QSPTs) have been the focus of significant theoretical and experimental studies [1]. QSPTs can be produced by varying a non-thermal control parameter at zero temperature. The neutron number, for example, can be one of these control parameters. In general, gradual transitions between various shapes in chains of isotopes (or isotones) predominate, although, in a few situations, rapid changes in ground-state characteristics are observed when moving from one isotope (or isotone) to its neighbor. A significant question emerges with QSPT from the perspective of experiments: Is it possible to find solvable models or symmetries that can be used to explain the structure at the phase transition point? Iachello [2, 3] proposed the notion of critical point symmetry (CPS), which describes nuclei at QSPT points between distinct dynamical symmetries. The CPSs give parameter-free predictions for the structural properties of nuclei at points of QSPTs. Typically, the CPSs were obtained using the Bohr Hamiltonian (BH) [4-6] with appropriate $\gamma$- and $\beta$-potentials depending on the physical circumstances under consideration. The second-order phase transition point between vibrational and $\gamma$-unstable shapes is the E(5) CPS [2]. In this case, the potential energy of the structure is independent of $\gamma$ and the characteristics of the nuclei at the CPS derived by utilizing an infinite-well potential for the $\beta$-part in BH [2]. The X(5) symmetry describes the first-order phase transition point between vibrational and rotational shapes. For this transitional region, the X(5) CPS [3] is an approximate solution using the sum of the harmonic oscillator potential (HOP) having a minimum at $\gamma = 0$ and an infinite-well potential. On the other hand, the Z(5) symmetry [7] is the phase transition point between prolate and oblate shapes. It has been experimentally identified in the $^{192}$Pt, $^{194}$Pt, and $^{196}$Pt isotopes [7]. The potential in the Z(5) model was assumed to be dependent on the collective variables $\beta$ and $\gamma$ and could be split into two parts, i.e., $u(\beta, \gamma) = u(\beta) + v(\gamma)$. Then the variables have been approximately separated for $\gamma = \pi/6$. In Z(5) model, it is reasonable to use an infinite well potential in the $\beta$-variable and HOP in the $\gamma$-variable [7].

The CPSs have paved the way for developing a wide range of new geometrical collective models by choosing different forms of the potential $u(\beta, \gamma)$ used in the BH and solving the relevant eigenvalue problem (analytically or approximately). Moreover, the exactly separable (ES) models provide a detailed description of nuclei near to or far away from the CPSs, for example the ES-X(5) [8] model. The most frequently used potentials include Davidson [9-13], Woods-Saxon [14], Sextic [15-24], Kratzer [25-27], Hulthen [28], and Pöschl-Teller [29].

Recently, using a conformable fractional version of the BH, we have developed a new category of CPSs called the conformable fractional $E^\alpha(5)$ CPS [30], where $\alpha$ is the order of the fractional derivative. The analytic formulae of the energy eigenvalues and eigenvectors of the conformable fractional Bohr Hamiltonian (CFBH) (with Kratzer potential in $\beta$ variable) were obtained [31]. In addition, the shape coexistence with mixing phenomena was studied in the $^{96}$Mo nucleus by using the CFBH [32]. The CFBH provides numerous models that are extremely similar to the original one, which enhances the probability of a high degree of accuracy with the experimental data, especially if $\alpha$ is close to 1.

The present study is intended to compare classical and conformable fractional analytical solutions of BH with three exponential-type potentials, namely Morse, Tietz-Hua, and multi-parameter exponential-type potential. Recently, these potentials have been attracting much interest in nuclear structure theory. For all values of angular momentum $L$, the potentials, such as a HOP, Coulomb, infinite-well, Kratzer, and Davidson, are known to be exactly solvable. On the other hand, it is also known that the Schrödinger equation cannot be solved exactly for Morse, Tietz-Hua, and multi-parameter exponential-type potential for $L \neq 0$ by using traditional methods like the super-symmetry [33] and the Nikiforov–Uvarov [34]. In this case, the effective potential combines the exponential and inverse square potentials, which cannot be solved analytically. As a result, an approximation procedure is required. The most commonly used and convenient one is the Pekeris approximation method [35]. The main idea of the Pekeris approximation is to expand approximately the centrifugal term using exponentials that are similar to those found in the rest of the potential, intending to absorb the centrifugal term into the potential.

Rather than the infinite-well potential used in the Z(5) CPS, the potential in $\beta$-part is assumed to be the Morse, Tietz-Hua, or multi-parameter exponential-type potential. At the same time, the $\gamma$-part is supposed to be the HOP, with a minimum at $\gamma = \pi/6$. We first succeeded in producing approximate solutions in a closed form of the CFBH for the Morse, Tietz-Hua, and multi-parameter exponential-type potentials with a centrifugal barrier. Following that, we fitted many triaxial nuclei, demonstrating the successes of these exponential-type potentials. Analytical formulae for the wave functions and energy spectra are determined by using the conformable fractional Nikiforov-Uvarov (CFNU) technique [36]. These solutions will be called the Z(5)-conformable fractional Morse (Z(5)-CFM), Z(5)-conformable fractional Tietz-Hua (Z(5)-CFTH), and Z(5)-conformable fractional exponential (Z(5)-CFE) models. The development of the energy spectrum as a function of the potential parameters is studied for the states in the ground, $\beta$, and $\gamma$ bands. The normalized $B$(E2) transition rates and the energy spectra are calculated and compared to theoretical results from different models and experimental findings. The theoretically calculated $B$(E2) transition rates and spectra are in good agreement with experimental data for the $^{114,116}$Pd, $^{126,128}$Xe, and $^{192,194}$Pt isotopes.

In section 2, the CFNU method is briefly reviewed. In section 3, the Bohr collective Hamiltonian and $B$(E2) transition rates for triaxial nuclei will be considered. In sections 4, 5, and 6, the closed formulae for the wave functions and energy spectra of Z(5)-CFM, Z(5)-CFTH, and Z(5)-CFE models are derived. Numerical results utilizing the CFBH will be reported in section 7, while section 8 includes an overview of the present results.

## 2. CFNU method

In recent decades, fractional calculus [37-39] has acquired growing attention from mathematicians and physicists due to its applications in applied mathematics and physics subjects, for instance, the investigation of vibrational and rotational spectra in nuclear structure. It is known that the first idea of a fractional derivative was proposed in 1695 by Leibnitz [40, 41]. On the other hand, one of the major challenges in the current development of fractional calculus is a proper approximation of fractional derivative and integral operators because an analytical solution for models including these operators is generally not available. Several types of fractional derivative definitions were suggested, such as Riemann-Liouville, Caputo, and Riesz. Most of the fractional derivatives are defined by fractional integrals. A new definition of a well-behaved fractional derivative, namely conformable fractional derivative, was introduced by Khalil et al. [42]. This definition is based on the limit definition of the derivative. The conformable fractional calculus obeys the basic properties of usual derivatives, such as quotient, product, and chain rules. These properties allow us to find analytical solutions for energy spectra, electromagnetic transition rates, and wave functions in many physical models. Several essential differential equations, especially in quantum mechanics, have been reformulated based on the definition of the conformable fractional derivative [36, 43, 44].

The conformable fractional derivative $\mathfrak{D}_s^\alpha f$ of order $\alpha$, for a given function $f : [s, \infty) \to \mathbb{R}$, was defined as

$$(\mathfrak{D}_s^\alpha f)(z) = \lim_{\mathfrak{h} \to 0} \frac{f(z + \mathfrak{h}(z-s)^{1-\alpha}) - f(z)}{\mathfrak{h}}, \quad (1)$$

for all $z > s$, $0 < \alpha \leq 1$. When $s = 0$, the corresponding fractional derivative operator becomes $\mathfrak{D}^\alpha$. The fundamental properties of the conformable fractional derivative and integral operators are as follows

(1) $\mathfrak{D}^\alpha(c_1 g_1 + c_2 g_2)(z) = c_1 \mathfrak{D}^\alpha g_1(z) + c_2 \mathfrak{D}^\alpha g_2(z)$, for all $c_1, c_2 \in \mathbb{R}$ and $g_1(z), g_2(z)$ are $\alpha$-differentiable for all positive $z$.
(2) $\mathfrak{D}^\alpha(g_1(z) g_2(z)) = g_1(z) \mathfrak{D}^\alpha(g_2(z)) + g_2(z) \mathfrak{D}^\alpha(g_1(z))$.
(3) $\mathfrak{D}^\alpha(g_1(z)/g_2(z)) = [g_2(z) \mathfrak{D}^\alpha(g_1(z)) - g_1(z) \mathfrak{D}^\alpha(g_2(z))]/g_2^2(z)$.

(4) $\mathfrak{D}^\alpha \omega = 0$, for constant function $\omega$.

(5) $\mathfrak{D}^\alpha z^t = t z^{t-\alpha}$, for all $t \in \mathbb{R}$.

(6) $I^\alpha g(z) = I^1\left(z^{\alpha-1} g(z)\right) = \int_s^z z^{\alpha-1} g(z) dz$, where $I^\alpha$ is the conformable fractional integral operator.

Recently, many considerable efforts have been devoted to studying the BH with different potentials to describe the structure of atomic nuclei. Analytical expressions for energy eigenvalues and the corresponding eigenfunctions of the Schrödinger equation were obtained by numerous traditional methods: the asymptotic iteration method [45], the supersymmetric approach [33], shape invariance method [46], the factorization method [47], and the ansatz method. Moreover, the Nikiforov-Uvarov method [34] was utilized to solve the hypergeometric-type equation. Due to the simplicity and high efficiency of the Nikiforov-Uvarov method, the Bohr equation can be solved systematically to obtain closed formulae for wave functions and energy spectra.

In the present section, the CFNU method in its complete conformable fractional form (the non-integer order form) is explained briefly to obtain the eigenvalue and the eigenfunction solutions of a fractional differential equation. The central equation of the CFNU method is

$$\mathfrak{D}^\alpha \mathfrak{D}^\alpha \psi(z) + \frac{\tilde{\tau}(z)}{\sigma(z)} \mathfrak{D}^\alpha \psi(z) + \frac{\tilde{\sigma}(z)}{\sigma^2(z)} \psi(z) = 0, \tag{2}$$

where $\sigma(z)$, $\tilde{\sigma}(z)$ and $\tilde{\tau}(z)$ are functions of at most $2\alpha$-th, $2\alpha$-th, and $\alpha$-th degree, correspondingly. Assuming

$$\psi(z) = \phi(z) \chi_{n,\alpha}(z), \tag{3}$$

(2) can be written as

$$\sigma(z) \mathfrak{D}^\alpha \mathfrak{D}^\alpha \chi_{n,\alpha}(z) + \tau(z) \mathfrak{D}^\alpha \chi_{n,\alpha}(z) + \lambda \chi_{n,\alpha}(z) = 0, \tag{4}$$

where $\phi(z)$ satisfies

$$\frac{\mathfrak{D}^\alpha \phi(z)}{\phi(z)} = \frac{\pi(z)}{\sigma(z)}, \tag{5}$$

and

$$\tau(z) = \tilde{\tau}(z) + 2\pi(z). \tag{6}$$

The derivative of the function $\tau(z)$ must be negative to obtain a physically acceptable solution. On the other hand, the functions $\chi_{n,\alpha}(z)$ are given by the Rodrigues-like formula,

$$\chi_{n,\alpha}(z) = \frac{\mathfrak{B}_n}{\rho(z)} (\mathfrak{D}^\alpha)^{(n)}\left(\sigma^n(z) \rho(z)\right), \quad n = 0,1,2,\ldots, \tag{7}$$

where $\mathfrak{B}_n$ is a normalization constant. The weight function $\rho(z)$ must satisfy the conformable fractional differential equations,

$$\mathfrak{D}^\alpha\left(\rho(z) \sigma(z)\right) = \rho(z) \tau(z). \tag{8}$$

For the CFNU method, the function $\pi(z)$ and the parameter $\lambda$ are essentially specified as

$$\pi(z) = \frac{\mathfrak{D}^\alpha \sigma(z) - \tilde{\tau}(z)}{2} \pm \sqrt{\left(\frac{\mathfrak{D}^\alpha \sigma(z) - \tilde{\tau}(z)}{2}\right)^2 - \tilde{\sigma}(z) + k\sigma(z)}, \tag{9}$$

and

$$\lambda = k + \mathfrak{D}^\alpha \pi(z). \tag{10}$$

To find the value of $k$, the expression under the square root of (9) must be the square of a function of a maximum degree $\alpha$. Hence, a new eigenvalue equation becomes

$$\lambda = \lambda_n = -\frac{n(n-1)}{2} \mathfrak{D}^\alpha \mathfrak{D}^\alpha \sigma(z) - n \mathfrak{D}^\alpha \tau(z). \tag{11}$$

## 3. Bohr collective Hamiltonian and $B(E2)$ transition rates

Our original collective BH [4-6] is

$$H = -\frac{\hbar^2}{2B} \left[\frac{1}{\beta^4} \frac{\partial}{\partial \beta} \beta^4 \frac{\partial}{\partial \beta} + \frac{1}{\beta^2 \sin 3\gamma} \frac{\partial}{\partial \gamma} \sin 3\gamma \frac{\partial}{\partial \gamma} - \frac{1}{4\beta^2} \sum_{\kappa=1,2,3} \frac{\hat{Q}_\kappa^2}{\sin^2(\gamma - 2\pi\kappa/3)}\right] + V(\beta,\gamma), \tag{12}$$

where $\beta$, and $\gamma$ represent the intrinsic collective coordinates, $\hat{Q}_\kappa$ are the angular momentum body-fixed components, and $B$ is the mass parameter. Whenever the potential has a minimum around $\gamma = \pi/6$, the sum term in (12), which contains the angular momentum components, transforms to

$$\sum_\kappa \frac{Q_\kappa^2}{\sin^2\left(\gamma - \frac{2}{3}\pi\kappa\right)} \approx 4(Q_1^2 + Q_2^2 + Q_3^2) - 3Q_1^2. \tag{13}$$

The separation of variables can be accomplished first by introducing reduced energies $\epsilon = \frac{2B}{\hbar^2}E$ and reduced potentials $u = \frac{2B}{\hbar^2}V$ and then assuming the reduced energies and potentials take the forms $\epsilon = \epsilon_\beta + \epsilon_\gamma$, and $u(\beta,\gamma) = u(\beta) + v(\gamma)$, respectively. Furthermore, suppose that the total wave function is

$$\Psi(\beta,\gamma,\theta_i) = \xi_{L,\varrho}(\beta)\eta(\gamma)\mathcal{D}^L_{M,\varrho}(\theta_i), \tag{14}$$

where $\theta_i (i = 1,2,3)$, and $\mathcal{D}^L_{M,\varrho}(\theta_i)$ are the Euler angles and Wigner functions of these angles. The symbols $M$ and $\varrho$ are the eigenvalues of the projections of angular momentum on the laboratory fixed $z$-axis, and the eigenvalues of the projections of angular momentum on the body-fixed $\hat{x}$-axis, respectively. The Schrödinger equation corresponding to the BH, (12), must be separable to

$$\left[-\frac{1}{\beta^4}\frac{\partial}{\partial \beta}\beta^4\frac{\partial}{\partial \beta} + \frac{1}{4\beta^2}(4L(L+1) - 3\varrho^2) + u(\beta)\right]\xi_{L,\varrho}(\beta) = \epsilon_\beta \xi_{L,\varrho}(\beta), \tag{15}$$

$$\left[-\frac{1}{\langle\beta^2\rangle \sin 3\gamma}\frac{\partial}{\partial\gamma}\sin 3\gamma\frac{\partial}{\partial\gamma} + u(\gamma)\right]\eta(\gamma) = \epsilon_\gamma \eta(\gamma), \tag{16}$$

where the first differential equation, (15), depends only on the $\beta$ collective variable, while the second equation, (16), depends on the $\gamma$-variable. Here, the quantum number $L$ represents the eigenvalues of the angular momentum operator, and $\langle\beta^2\rangle$ is the average of $\beta^2$ over the wave function $\xi_{L,\varrho}(\beta)$.

Equation (16) has a simple form, so, we begin with its solution. For the triaxial nuclei, the potential $u(\gamma)$ is considered to be a HOP in the form $u(\gamma) = c\bar{\gamma}^2/2$, with $\bar{\gamma} = \gamma - \pi/6$. In this case, by inserting $u(\gamma)$ into differential equation (16), we obtain the wave function, for $\gamma = \pi/6$, in terms of Hermite polynomials; for more details, see Ref. [7], and the $\gamma$-part of the eigenvalue solutions becomes $\varepsilon_{\bar{\gamma}} = \sqrt{2c/\langle\beta^2\rangle}(n_{\bar{\gamma}} + 1/2)$, for $n_{\bar{\gamma}} = 0, 1, 2, ...$.

Applying the known relation $\varrho = L - n_w$, where $n_w$ is the wobbling quantum number [6, 48], (15) becomes

$$\left[-\frac{1}{\beta^4}\frac{\partial}{\partial\beta}\beta^4\frac{\partial}{\partial\beta} + \frac{1}{4\beta^2}(L(L+4) + 3n_w(2L - n_w)) + u(\beta)\right]\xi_{n_w,L}(\beta) = \epsilon_\beta \xi_{n_w,L}(\beta). \tag{17}$$

By setting $\bar{\xi}_{n_w,L}(\beta) = \beta^2 \xi_{n_w,L}(\beta)$ in (17), one obtains

$$\left[-\frac{d^2}{d\beta^2} + \frac{\nu+2}{\beta^2} + u(\beta) - \epsilon_\beta\right]\bar{\xi}_{n_w,L}(\beta) = 0, \tag{18}$$

where

$$\nu = \frac{L(L+4) + 3n_w(2L-n_w)}{4}. \tag{19}$$

After the determination of the wave function by using the strategy followed in the CFNU method, it is straightforward to calculate the $B(E2)$ transition rates by utilizing these wave functions. The $B(E2)$ transition rates are computed using matrix elements of the transition operators. In this aspect, the quadrupole operator is extremely useful [7],

$$T^{(E2)}_M = t\beta\left[\mathcal{D}^{(2)}_{M,0}(\theta_i)\cos\left(\gamma - \frac{2\pi}{3}\right) + \frac{1}{\sqrt{2}}\left(\mathcal{D}^{(2)}_{M,2}(\theta_i) + \mathcal{D}^{(2)}_{M,-2}(\theta_i)\right)\sin\left(\gamma - \frac{2\pi}{3}\right)\right], \tag{20}$$

where $t$ is a scaling factor.

In the case of $\gamma \simeq \pi/6$, the operator $T^{(E2)}_M$ transforms to

$$T^{(E2)}_M = -\frac{1}{\sqrt{2}}t\beta\left(\mathcal{D}^{(2)}_{M,2}(\theta_i) + \mathcal{D}^{(2)}_{M,-2}(\theta_i)\right). \tag{21}$$

Hence, the $B(E2)$ transition rates can be written as

$$B(E2; L_i\varrho_i \to L_f\varrho_f) = \frac{5}{16\pi}\frac{|\langle L_f\varrho_f\|T^{(E2)}\|L_i\varrho_i\rangle|^2}{(2L_i+1)}, \tag{22}$$

where the reduced matrix element is computed using the Wigner-Eckart theorem,

$$\langle L_f\varrho_f|T^{(E2)}_M|L_i\varrho_i\rangle = \frac{(L_i 2L_f|\varrho_i M\varrho_f)}{\sqrt{2L_f+1}}\langle L_f\varrho_f\|T^{(E2)}\|L_i\varrho_i\rangle. \tag{23}$$

In calculating the matrix elements of (23), the integral over $\beta$ will be expressed in the form

$$I_\beta(n_i, L_i, \varrho_i, n_f, L_f, \varrho_f) = \int_0^\infty \beta \xi_{n_i,\varrho_i,L_i}(\beta)\xi_{n_f,\varrho_f,L_f}(\beta)\beta^4 d\beta. \tag{24}$$

Accordingly, the final formula of the $B(E2)$ is given by

$$B(E2; L_i\varrho_i \to L_f\varrho_f)$$
$$= \frac{5}{16\pi}\frac{t^2}{2}\frac{1}{(1+\delta_{\varrho_i,0})(1+\delta_{\varrho_f,0})}\left[(L_i 2L_f|\varrho_i 2\varrho_f) + (L_i 2L_f|\varrho_i - 2\varrho_f)\right.$$
$$\left. + (-)^{L_i}(L_i 2L_f|-\varrho_i 2\varrho_f)\right]^2 I_\beta^2(n_i, L_i, \varrho_i, n_f, L_f, \varrho_f). \tag{25}$$

It has to be stressed that, in (25), the Clebsch–Gordan coefficients yield a $\Delta\varrho = \pm 2$ selection rule.

## 4. Morse potential

The Morse potential has attracted a lot of attention over the years, and it is shown to be one of the most valuable models for describing nuclei in the $\gamma$-unstable and rotational with $\gamma \approx 0$ cases. The Morse potential was also employed [49, 50] to address the overprediction of the energy spacing problem inside the $\beta$ band of the X(5) CPS and other related solutions. Moreover, this potential was used in the $\beta$-part of the collective BH for triaxial nuclei [51].

The Morse potential in the $\beta$ variable is given by [52]
$$u(\beta) = u_0\left(e^{-2a(\beta-\beta_e)} - 2e^{-a(\beta-\beta_e)}\right). \tag{26}$$
The quantities $a$ (shape parameter), that controls the width of the potential, and $\beta_e$ (position of the minimum) entering the definition (26), allow for some adjustments, and make the choice of Morse potential reasonably flexible. The overall parameter $u_0$ is set equal to unity, without altering the solution, if ratios of energies are employed, as in the present work. This potential becomes flat on the right-hand side, i.e., for large $\beta$. Equation (18) can be rewritten with Morse potential as
$$\left[\frac{d^2}{d\beta^2} + \epsilon_M - \frac{\nu+2}{\beta^2} - e^{-2a(\beta-\beta_e)} + 2e^{-a(\beta-\beta_e)}\right]\bar{\xi}_{n_w,L}(\beta) = 0. \tag{27}$$

Defining
$$x = \frac{\beta - \beta_e}{\beta_e}, \qquad \delta = a\beta_e, \qquad \varepsilon_M = \beta_e^2 \epsilon_M, \qquad \mu = \nu + 2, \tag{28}$$

so (27) becomes
$$\left[\frac{d^2}{dx^2} + \varepsilon_M - \frac{\mu}{(x+1)^2} - \beta_e^2 e^{-2\delta x} + 2\beta_e^2 e^{-\delta x}\right]\bar{\xi}_{n_w,L}(x) = 0. \tag{29}$$

The Bohr equation cannot be solved exactly for the exponential potentials such as the Morse one by using the standard methods. Consequently, it is convenient to use the idea of the Pekeris approximation. In the Pekeris approximation [35], the centrifugal term is expanded in a series of exponentials depending on the variable $x$, maintaining terms up to second-order. In this case, the effective $L$-dependent potential retains the same shape as the potential with $L = 0$ [53]. Expanding $u_L(x)$ into binomial series, we have
$$u_L(x) = \frac{\mu}{(x+1)^2} = \mu(1 - 2x + 3x^2 - 4x^3 + \cdots). \tag{30}$$

As it was mentioned above, the basic idea of the Pekeris approximation is to rewrite approximately the centrifugal term using exponentials resembling the ones appearing in the rest of the potential, with the aim of being able to 'absorb' the centrifugal term into the potential. In the exponential form, $u_L(x)$ can be expressed as
$$\tilde{u}_L(x) \approx \mu\left(c_0 + c_1 e^{-\delta x} + c_2 e^{-2\delta x} + \cdots\right). \tag{31}$$
In order to determine the constants $c_0$, $c_1$, and $c_2$, (31) is expanded in a series of powers of $x$ around $x = 0$ to give
$$\tilde{u}_L(x) \approx \mu\left((c_0 + c_1 + c_2) - (c_1 + 2c_2)\delta x + \left(\frac{1}{2}c_1 + 2c_2\right)\delta^2 x^2 - (c_1 + 8c_2)\frac{\delta^3 x^3}{6} + \cdots\right). \tag{32}$$

Combining equal powers of (32) with (30), we obtain:
$$c_0 = 1 - \frac{3}{\delta} + \frac{3}{\delta^2}, \qquad c_1 = \frac{4}{\delta} - \frac{6}{\delta^2}, \qquad c_2 = \frac{3}{\delta^2} - \frac{1}{\delta}. \tag{33}$$

Substituting (31) into (29), we have
$$\left(\frac{d^2}{dx^2} + (\varepsilon_M - \mu c_0) + (2\beta_e^2 - \mu c_1)e^{-\delta x} - (\mu c_2 + \beta_e^2)e^{-2\delta x}\right)\bar{\xi}_{n_w,L}(x) = 0. \tag{34}$$

By setting [54]
$$\varepsilon_M - \mu c_0 = -\rho^2, \qquad 2\beta_e^2 - \mu c_1 = \gamma_1^2, \qquad \mu c_2 + \beta_e^2 = \gamma_2^2, \tag{35}$$
equation (34) transforms to
$$\left[\frac{d^2}{dx^2} - \rho^2 + \gamma_1^2 e^{-\delta x} - \gamma_2^2 e^{-2\delta x}\right]\bar{\xi}_{n_w,L}(x) = 0. \tag{36}$$

Changing the variable in (36) by introducing $y = e^{-\delta x}$ and applying simplifications lead to a differential equation
$$\left[\frac{d^2}{dy^2} + \frac{1}{y}\frac{d}{dy} + \frac{(-\gamma_2^2 y^2 + \gamma_1^2 y - \rho^2)}{\delta^2 y^2}\right]\bar{\xi}_{n_w,L}(y) = 0. \tag{37}$$

Equation (37) can be represented in fractional form by replacing the differentiation orders and function degrees in the coefficients with fractional orders [36]. Hence, the conformable fractional version of (37) is,

$$\left[\mathfrak{D}^\alpha\mathfrak{D}^\alpha + \frac{1}{y^\alpha}\mathfrak{D}^\alpha + \frac{(-\gamma_2^2 y^{2\alpha} + \gamma_1^2 y^\alpha - \rho^2)}{\delta^2 y^{2\alpha}}\right]\bar{\xi}_{n_w,L}(y) = 0. \tag{38}$$

The functions $\tilde{\tau}(y)$, $\sigma(y)$, and $\tilde{\sigma}(y)$ are extracted by comparing (38) with (2)

$$\tilde{\tau}(y) = 1, \qquad \sigma(y) = y^\alpha, \qquad \tilde{\sigma}(y) = \frac{1}{\delta^2}(-\gamma_2^2 y^{2\alpha} + \gamma_1^2 y^\alpha - \rho^2). \tag{39}$$

Solutions of (9) for the functions $\pi(y)$ together with the functions defined in (39), are

$$\pi(y) = \frac{\alpha - 1}{2} \pm \sqrt{\frac{\gamma_2^2}{\delta^2}y^{2\alpha} + \left(k - \frac{\gamma_1^2}{\delta^2}\right)y^\alpha + \left(\frac{\alpha-1}{2}\right)^2 + \frac{\rho^2}{\delta^2}}. \tag{40}$$

Taking the discriminant of the quadratic equation within the square root sign of (40) to be zero, double roots of $k$ are obtained;

$$k_\pm = \frac{\gamma_1^2}{\delta^2} \pm \frac{\gamma_2}{\delta^2}\Omega, \qquad \Omega = \sqrt{(\alpha-1)^2\delta^2 + 4\rho^2}. \tag{41}$$

With the substitution of (41), (40) for $\pi(y)$ has the following four possible forms,

$$\pi(y) = \frac{1}{2}(\alpha - 1) \pm \begin{cases} \frac{\gamma_2}{\delta}y^\alpha + \frac{\Omega}{2\delta} & \text{for } k_+ = \frac{\gamma_1^2}{\delta^2} + \frac{\gamma_2}{\delta^2}\Omega, \\ \frac{\gamma_2}{\delta}y^\alpha - \frac{\Omega}{2\delta} & \text{for } k_- = \frac{\gamma_1^2}{\delta^2} - \frac{\gamma_2}{\delta^2}\Omega. \end{cases} \tag{42}$$

The function $\tau(y)$ satisfy the requirement that

$$\tau(y) = \alpha + \frac{\Omega}{\delta} - \frac{2\gamma_2}{\delta}y^\alpha, \qquad \mathfrak{D}^\alpha\tau(y) = -\frac{2\gamma_2}{\delta}\alpha < 0, \tag{43}$$

leads to a physically acceptable solution, and the function, $\pi(y)$, in this case, is selected as

$$\pi(y) = \frac{1}{2}(\alpha - 1) - \frac{\gamma_2}{\delta}y^\alpha + \frac{\Omega}{2\delta}, \tag{44}$$

for $k_- = \frac{\gamma_1^2}{\delta^2} - \frac{\gamma_2}{\delta^2}\Omega$. To get the energy eigenvalue solution, the expressions of $\lambda$ and $\lambda_n$ must be determined by using (10) and (11), respectively,

$$\lambda = \frac{\gamma_1^2}{\delta^2} - \frac{\gamma_2}{\delta^2}\Omega - \frac{\alpha\gamma_2}{\delta}, \qquad \lambda_n = \frac{2n\alpha\gamma_2}{\delta}. \tag{45}$$

The two last expressions of $\lambda$ and $\lambda_n$ must be equated such that this equality yields the following expression of the energy

$$\epsilon_M = \frac{1}{4\beta_e^2}(\alpha-1)^2\delta^2 + \frac{\mu c_0}{\beta_e^2} - \frac{1}{4\beta_e^2}\left(\frac{\gamma_1^2}{\gamma_2} - (2n+1)\alpha\delta\right)^2. \tag{46}$$

The evolution of the Z(5)-CFM energy levels of the ground, $\beta$, and $\gamma$ bands as a function of the potential parameters and angular momenta are shown in Figs. 1 and 2, respectively.

In case $\alpha = 1$, the expression of the energy eigenvalue is of the form

$$\epsilon_M = \frac{\mu c_0}{\beta_e^2} - \left(\frac{\gamma_1^2}{2\beta_e\gamma_2} - \left(n + \frac{1}{2}\right)\frac{\delta}{\beta_e}\right)^2. \tag{47}$$

Equation (47) is in complete concurrence with (24) of Ref. [49].

To calculate the corresponding wave function, using the strategy of the CFNU method, the function $\bar{\xi}_{n,n_w,L}(\beta)$ is obtained by defining

$$\bar{\xi}_{n,n_w,L}(\beta) = \phi(\beta)\chi_{n,\alpha}(\beta). \tag{48}$$

After imposing the expressions $\sigma(\beta)$ and $\pi(\beta)$ into (5), and applying the properties of the conformable fractional integral [42], the function $\phi(\beta)$, i.e. first part of the $\bar{\xi}_{n,n_w,L}(\beta)$, can be determined as

$$\phi(\beta) = e^{-\frac{1}{2}((\alpha-1)\delta+\Omega)\left(\frac{\beta-\beta_e}{\beta_e}\right) - \frac{\gamma_2}{\delta\alpha}e^{-\alpha\delta\left(\frac{\beta-\beta_e}{\beta_e}\right)}}. \tag{49}$$

Following that, the weight function $\rho(\beta)$ should be determined by using (8),

$$\rho(\beta) = e^{-\Omega\left(\frac{\beta-\beta_e}{\beta_e}\right) - \frac{2\gamma_2}{\delta\alpha}e^{-\alpha\delta\left(\frac{\beta-\beta_e}{\beta_e}\right)}}, \tag{50}$$

and the function $\chi_{n,\alpha}(\beta)$ is now obtained by substituting the function $\rho(\beta)$ into (7),

$$\chi_{n,\alpha}(\beta) = \mathfrak{B}_{n,n_w,L}e^{\Omega\left(\frac{\beta-\beta_e}{\beta_e}\right) + \frac{2\gamma_2}{\delta\alpha}e^{-\alpha\delta\left(\frac{\beta-\beta_e}{\beta_e}\right)}}(\mathfrak{D}^\alpha)^{(n)}\left(e^{-(n\alpha\delta+\Omega)\left(\frac{\beta-\beta_e}{\beta_e}\right) - \frac{2\gamma_2}{\delta\alpha}e^{-\alpha\delta\left(\frac{\beta-\beta_e}{\beta_e}\right)}}\right), \quad n = 0,1,2,..., \tag{51}$$

in which $\mathfrak{B}_{n,n_w,L}$ are the normalization constants. For the sake of completeness, using (48), (49), and (51) in addition to the transformation $\bar{\xi}_{n,n_w,L}(\beta) = \beta^2\xi_{n,n_w,L}(\beta)$, we give the physically acceptable wave function $\xi_{n,n_w,L}(\beta)$ in the form

$$\xi_{n,n_w,L}(\beta) = \mathfrak{B}_{n,n_w,L}\beta^{-2}e^{-\frac{1}{2}((\alpha-1)\delta-\Omega)\left(\frac{\beta-\beta_e}{\beta_e}\right)+\frac{\gamma_2}{\delta\alpha}e^{-\alpha\delta\left(\frac{\beta-\beta_e}{\beta_e}\right)}}(\mathfrak{D}^\alpha)^{(n)}\left(e^{-(n\alpha\delta+\Omega)\left(\frac{\beta-\beta_e}{\beta_e}\right)-\frac{2\gamma_2}{\delta\alpha}e^{-\alpha\delta\left(\frac{\beta-\beta_e}{\beta_e}\right)}}\right), \quad n = 0,1,2,\ldots, \tag{52}$$

where the constants $\mathfrak{B}_{n,n_w,L}$ are determined based on the formula, $\int_0^\infty \beta^4 \xi_{n,n_w,L}^2(\beta)d\beta = 1$. When $n = 0$, the expression of the wave function has the form

$$\xi_{0,n_w,L}(\beta) = \mathfrak{B}_{0,n_w,L}\beta^{-2}e^{-\frac{1}{2}((\alpha-1)\delta+\Omega)\left(\frac{\beta-\beta_e}{\beta_e}\right)-\frac{\gamma_2}{\delta\alpha}e^{-\alpha\delta\left(\frac{\beta-\beta_e}{\beta_e}\right)}}, \tag{53}$$

and the corresponding normalization constant becomes

$$\mathfrak{B}_{0,n_w,L} = \frac{1}{\sqrt{\int_0^\infty e^{-((\alpha-1)\delta+\Omega)\left(\frac{\beta-\beta_e}{\beta_e}\right)-\frac{2\gamma_2}{\delta\alpha}e^{-\alpha\delta\left(\frac{\beta-\beta_e}{\beta_e}\right)}}d\beta}}. \tag{54}$$

## 5. Tietz-Hua potential

The Tietz-Hua potential is one of the most accurate analytical models for describing the vibrational energy spectra of diatomic molecules [55, 56]. It is significantly more realistic than the Morse potential in explaining molecular dynamics at high rotational and vibrational quantum numbers [57]. Recently [58], in nuclear structure theory, the solutions of the BH with the Tietz-Hua potential were given, which was utilized to describe the $\beta$-part of the potential and the HOP for the $\gamma$-part. The computations were performed for energy levels and $B(E2)$ transition rates for $\gamma$-unstable and axially symmetric deformed nuclei. In this section, new analytical expressions for the energy spectra and the corresponding wavefunctions for triaxial nuclei are derived with the Tietz-Hua potential in the conformable fractional framework.

The Tietz-Hua potential is given in the following form

$$u(\beta) = D_e\left[\frac{1-e^{-b_h(\beta-\beta_e)}}{1-c_h e^{-b_h(\beta-\beta_e)}}\right]^2, \tag{55}$$

where $b_h$, $c_h$, $D_e$ (the potential depth) and $\beta_e$ (the potential minimum) are free parameters. In numerical calculations, the parameter $D_e$ will be assumed to be unity for simplicity. So (18) becomes

$$\left[\frac{d^2}{d\beta^2} + \epsilon_{TH} - V_l(\beta) - \left[\frac{1-e^{-b_h(\beta-\beta_e)}}{1-c_h e^{-b_h(\beta-\beta_e)}}\right]^2\right]\bar{\xi}_{n_w,L}(\beta) = 0, \tag{56}$$

where $V_l(\beta) = \frac{v+2}{\beta^2}$. Equation (56) cannot be solved analytically for $l \neq 0$ due to the centrifugal term. As a result, once more, we will utilize the Pekeris approximation to handle the centrifugal term. In this approximation, for a small $\beta$, $V_l(\beta)$ could be expanded around $\beta = \beta_e$ in a series of powers of $x = \frac{(\beta-\beta_e)}{\beta_e}$ as

$$V_l(\beta) = \frac{v+2}{\beta^2} = \frac{v+2}{\beta_e^2}(1 - 2x + 3x^2 - 4x^3 + \cdots). \tag{57}$$

Expansion terms should only be kept up to the second order. In the Pekeris approximation [35], using the following form of the potential

$$\tilde{V}_l(\beta) \approx \frac{v+2}{\beta_e^2}\left(D_0 + D_1\frac{e^{-\delta x}}{1-c_h e^{-\delta x}} + D_2\frac{e^{-2\delta x}}{(1-c_h e^{-\delta x})^2}\right), \tag{58}$$

with

$$D_0 = 1 - \frac{1}{\delta}(1-c_h)(3+c_h) + \frac{3}{\delta^2}(1-c_h)^2, \quad \lim_{c_h \to 0} D_0 = 1 - \frac{3}{\delta} + \frac{3}{\delta^2}, \tag{59}$$

$$D_1 = \frac{2}{\delta}(1-c_h)^2(2+c_h) - \frac{6}{\delta^2}(1-c_h)^3, \quad \lim_{c_h \to 0} D_1 = \frac{4}{\delta} + \frac{6}{\delta^2}, \tag{60}$$

$$D_2 = -\frac{1}{\delta}(1-c_h)^3(1+c_h) + \frac{3}{\delta^2}(1-c_h)^4, \quad \lim_{c_h \to 0} D_2 = -\frac{1}{\delta} + \frac{3}{\delta^2}, \tag{61}$$

then (56) becomes,

$$\left[\frac{d^2}{d\beta^2} + \epsilon_{TH} - \frac{v+2}{\beta_e^2}\left(D_0 + D_1\frac{e^{-\delta x}}{1-c_h e^{-\delta x}} + D_2\frac{e^{-2\delta x}}{(1-c_h e^{-\delta x})^2}\right) - \left(\frac{1-e^{-\delta\frac{(\beta-\beta_e)}{\beta_e}}}{1-c_h e^{-\delta\frac{(\beta-\beta_e)}{\beta_e}}}\right)^2\right]\bar{\xi}_{n_w,L}(\beta) = 0, \tag{62}$$

where $\delta = b_h \beta_e$. Using the new variable $z = c_h e^{-\delta \frac{(\beta - \beta_e)}{\beta_e}}$, one gets

$$\left( \frac{d^2}{dz^2} + \frac{(1-z)}{(z-z^2)} \frac{d}{dz} + \frac{-\eta_1 z^2 + \eta_2 z - \eta_3}{(z-z^2)^2} \right) \bar{\xi}_{n_w,L}(z) = 0, \tag{63}$$

where

$$\eta_1 = \frac{\nu+2}{\delta^2} \left( D_0 - \frac{D_1}{c_h} + \frac{D_2}{c_h^2} \right) - \frac{\beta_e^2}{\delta^2} \left( \epsilon_{TH} - \frac{1}{c_h^2} \right), \tag{64}$$

$$\eta_2 = \frac{\nu+2}{\delta^2} \left( 2D_0 - \frac{D_1}{c_h} \right) - 2 \frac{\beta_e^2}{\delta^2} \left( \epsilon_{TH} - \frac{1}{c_h} \right), \tag{65}$$

$$\eta_3 = \frac{\nu+2}{\delta^2} D_0 - \frac{\beta_e^2}{\delta^2} (\epsilon_{TH} - 1). \tag{66}$$

By using the same procedures applied previously for the Morse potential, the fractional form of (63) is

$$\mathfrak{D}^\alpha \mathfrak{D}^\alpha \bar{\xi}_{n_w,L}(z) + \frac{(1-z^\alpha)}{(z^\alpha - z^{2\alpha})} \mathfrak{D}^\alpha \bar{\xi}_{n_w,L}(z) + \frac{-\eta_1 z^{2\alpha} + \eta_2 z^\alpha - \eta_3}{(z^\alpha - z^{2\alpha})^2} \bar{\xi}_{n_w,L}(z) = 0. \tag{67}$$

To identify coefficients in the fundamental equation of the CFNU method, (67) is compared with (2):

$$\tilde{\tau}(z) = 1 - z^\alpha, \qquad \sigma(z) = z^\alpha - z^{2\alpha}, \qquad \tilde{\sigma}(z) = -\eta_1 z^{2\alpha} + \eta_2 z^\alpha - \eta_3. \tag{68}$$

Substituting these functions in (9), the functions $\pi(z)$ can be derived. We get,

$$\pi(z) = \frac{(\alpha - 1) + (1 - 2\alpha) z^\alpha}{2} \pm \sqrt{\left( \zeta_1 - \frac{\beta_e^2}{\delta^2} \epsilon_{TH} - k \right) z^{2\alpha} - \left( \zeta_2 - 2 \frac{\beta_e^2}{\delta^2} \epsilon_{TH} - k \right) z^\alpha + \left( \zeta_3 - \frac{\beta_e^2}{\delta^2} \epsilon_{TH} \right)}, \tag{69}$$

where $\zeta_1 = \frac{(1-2\alpha)^2}{4} + \frac{\nu+2}{\delta^2} \left( D_0 - \frac{D_1}{c_h} + \frac{D_2}{c_h^2} \right) + \frac{\beta_e^2}{\delta^2} \left( \frac{1}{c_h^2} \right)$, $\zeta_2 = -\frac{2(\alpha-1)(1-2\alpha)}{4} + \frac{\nu+2}{\delta^2} \left( 2D_0 - \frac{D_1}{c_h} \right) + 2 \frac{\beta_e^2}{\delta^2} \left( \frac{1}{c_h} \right)$, and $\zeta_3 = \frac{(\alpha-1)^2}{4} + \frac{\nu+2}{\delta^2} D_0 + \frac{\beta_e^2}{\delta^2}$. According to the CFNU method, the term under the square root must be square of a function of a maximum degree $\alpha$. As a result, we have two possible roots of $k$;

$$k_\pm = \zeta_2 - 2\zeta_3 \pm 2\varpi\vartheta, \tag{70}$$

where $\vartheta = \sqrt{\zeta_3 - \frac{\beta_e^2}{\delta^2} \epsilon_{TH}}$, and $\varpi = \sqrt{\zeta_1 - \zeta_2 + \zeta_3}$. The possible forms of the functions $\pi(z)$, for each $k$, are as follow

$$\pi(z) = \frac{(\alpha - 1) + (1 - 2\alpha) z^\alpha}{2} \pm \begin{cases} (\vartheta - \varpi) z^\alpha - \vartheta, & \text{for } k_+ = \zeta_2 - 2\zeta_3 + 2\varpi\vartheta, \\ (\vartheta + \varpi) z^\alpha - \vartheta, & \text{for } k_- = \zeta_2 - 2\zeta_3 - 2\varpi\vartheta. \end{cases} \tag{71}$$

To get a physical eigenvalue, function $\pi(z)$ is chosen such that function $\tau(z)$ will have a negative derivative [34]. Consequently, $\pi(z)$ is chosen as

$$\pi(z) = \frac{(\alpha - 1) + (1 - 2\alpha) z^\alpha}{2} - (\vartheta + \varpi) z^\alpha + \vartheta, \tag{72}$$

for $k_- = \zeta_2 - 2\zeta_3 - 2\varpi\vartheta$. As a consequence, $\tau(z)$ becomes

$$\tau(z) = -2(\alpha + \vartheta + \varpi) z^\alpha + (\alpha + 2\vartheta). \tag{73}$$

Using (10) and (11), we find out the following expressions of $\lambda$ and $\lambda_n$ respectively:

$$\lambda = \zeta_2 - 2\zeta_3 - 2\varpi\vartheta - \alpha \left( \alpha - \frac{1}{2} \right) - \alpha(\vartheta + \varpi), \tag{74}$$

$$\lambda_n = n(n-1)\alpha^2 + 2n\alpha(\alpha + \vartheta + \varpi). \tag{75}$$

The energy eigenvalues are calculated using the relation $\lambda = \lambda_n$. Eventually, the analytical expression of the energy levels generated by the potential in (55) is given by

$$\epsilon_{TH} = 1 + \frac{(\alpha-1)^2 \delta^2}{4\beta_e^2} + \frac{(\nu+2)}{\beta_e^2} D_0 - \frac{\delta^2}{\beta_e^2} \left[ \frac{n(n+1)\alpha^2 + \alpha \left( \alpha - \frac{1}{2} \right) + (1+2n)\alpha \Upsilon + 2\zeta_3 - \zeta_2}{(1+2n)\alpha + 2\Upsilon} \right]^2, \tag{76}$$

with $\Upsilon = \sqrt{\frac{\alpha^2}{4} + \frac{\nu+2}{\delta^2} \left( \frac{D_2}{c_h^2} \right) + \frac{\beta_e^2}{\delta^2} \left( \frac{c_h - 1}{c_h} \right)^2}$. The evolution of the Z(5)-CFTH energy levels of the ground, $\beta$, and $\gamma$ bands as a function of the potential parameters and angular momenta are shown in Figs. 3 and 4, respectively.

Moreover, in the case of $\alpha = 1$, one has,

$$\epsilon_{TH} = 1 + \frac{(\nu+2)}{\beta_e^2}D_0 - \frac{\delta^2}{\beta_e^2}\left[\frac{n(n+1) + \frac{1}{2} + (1+2n)\sqrt{\frac{1}{4} + \frac{\nu+2}{\delta^2}\left(\frac{D_2}{c_h^2}\right) + \frac{\beta_e^2}{\delta^2}\left(\frac{c_h-1}{c_h}\right)^2} + \frac{(\nu+2)}{\delta^2}\left(\frac{D_1}{c_h}\right) + 2\frac{\beta_e^2}{\delta^2}\left(\frac{c_h-1}{c_h}\right)}{(1+2n) + 2\sqrt{\frac{1}{4} + \frac{\nu+2}{\delta^2}\left(\frac{D_2}{c_h^2}\right) + \frac{\beta_e^2}{\delta^2}\left(\frac{c_h-1}{c_h}\right)^2}}\right]^2. \tag{77}$$

Equation (77) is in full concurrence with (20) of Ref. [59].

In the CFNU method, the wave function is written as $\bar{\xi}_{n,n_w,L}(\beta) = \phi(\beta)\chi_{n,\alpha}(\beta)$. The function $\phi(\beta)$ is derived from (5),

$$\phi(\beta) = c_h^{\left(\frac{(\alpha-1)}{2}+\vartheta\right)} e^{-\delta\left(\frac{(\alpha-1)}{2}+\vartheta\right)\frac{(\beta-\beta_e)}{\beta_e}} \left(1 - c_h^\alpha e^{-\frac{\alpha\delta(\beta-\beta_e)}{\beta_e}}\right)^{\left(\frac{1}{2}+\frac{\varpi}{\alpha}\right)}. \tag{78}$$

However, to derive the function $\chi_{n,\alpha}(\beta)$, the weight function $\rho(\beta)$ must be determined first. Using (8), the weight function can be expressed as

$$\rho(\beta) = c_h^{2\vartheta} e^{-\frac{2\delta\vartheta(\beta-\beta_e)}{\beta_e}} \left(1 - c_h^\alpha e^{-\frac{\delta\alpha(\beta-\beta_e)}{\beta_e}}\right)^{\frac{2\varpi}{\alpha}}. \tag{79}$$

The function $\chi_{n,\alpha}(\beta)$ is obtained by substituting $\rho(\beta)$ into (7),

$$\chi_{n,\alpha}(\beta) = \mathfrak{B}_{n,n_w,L} c_h^{-2\vartheta} e^{\frac{2\delta\vartheta(\beta-\beta_e)}{\beta_e}} \left(1 - c_h^\alpha e^{-\frac{\delta\alpha(\beta-\beta_e)}{\beta_e}}\right)^{-\frac{2\varpi}{\alpha}}$$
$$\times (\mathfrak{D}^\alpha)^{(n)} \left(c_h^{n\alpha+2\vartheta} e^{-\frac{\delta(n\alpha+2\vartheta)(\beta-\beta_e)}{\beta_e}} \left(1 - c_h^\alpha e^{-\frac{\delta\alpha(\beta-\beta_e)}{\beta_e}}\right)^{n+\frac{2\varpi}{\alpha}}\right), \quad n = 0,1,2,..., \tag{80}$$

where $\mathfrak{B}_{n,n_w,L}$ are the normalization constants. From (78) and (80), the complete wave function $\xi_{n,n_w,L}(\beta)$ is simply determined by

$$\xi_{n,n_w,L}(\beta) = \mathfrak{B}_{n,n_w,L} \beta^{-2} c_h^{\left(\frac{(\alpha-1)}{2}-\vartheta\right)} e^{-\delta\left(\frac{(\alpha-1)}{2}-\vartheta\right)\frac{(\beta-\beta_e)}{\beta_e}} \left(1 - c_h^\alpha e^{-\frac{\alpha\delta(\beta-\beta_e)}{\beta_e}}\right)^{\left(\frac{1}{2}-\frac{\varpi}{\alpha}\right)}$$
$$\times (\mathfrak{D}^\alpha)^{(n)} \left(c_h^{n\alpha+2\vartheta} e^{-\frac{\delta(n\alpha+2\vartheta)(\beta-\beta_e)}{\beta_e}} \left(1 - c_h^\alpha e^{-\frac{\delta\alpha(\beta-\beta_e)}{\beta_e}}\right)^{n+\frac{2\varpi}{\alpha}}\right), \quad n = 0,1,2,..., \tag{81}$$

and the normalization condition $\int_0^\infty \beta^4 \xi_{n,n_w,L}^2(\beta) d\beta = 1$ is used to evaluate the constants $\mathfrak{B}_{n,n_w,L}$. In the case of $n = 0$, the wave function reads

$$\xi_{0,n_w,L}(\beta) = \mathfrak{B}_{0,n_w,L} \beta^{-2} c_h^{\left(\frac{(\alpha-1)}{2}+\vartheta\right)} e^{-\delta\left(\frac{(\alpha-1)}{2}+\vartheta\right)\frac{(\beta-\beta_e)}{\beta_e}} \left(1 - c_h^\alpha e^{-\frac{\alpha\delta(\beta-\beta_e)}{\beta_e}}\right)^{\left(\frac{1}{2}+\frac{\varpi}{\alpha}\right)}, \tag{82}$$

where

$$\mathfrak{B}_{0,n_w,L} = \frac{1}{\sqrt{\int_0^\infty c_h^{((\alpha-1)+2\vartheta)} e^{-\delta((\alpha-1)+2\vartheta)\frac{(\beta-\beta_e)}{\beta_e}} \left(1 - c_h^\alpha e^{-\frac{\alpha\delta(\beta-\beta_e)}{\beta_e}}\right)^{\left(1+\frac{2\varpi}{\alpha}\right)} d\beta}}. \tag{83}$$

## 6. The multi-parameter exponential-type potential

Many exactly solvable potentials are exponential functions of the spatial coordinate. These exponential potentials are extensively utilized in various physics areas. An ideal example is the Hulthén potential [60], which is used in atomic physics [61], solid-state physics [62, 63], chemical physics [64], and nuclear physics [65]. Many of the exactly solvable exponential-type potentials are reducible potentials. For instance, by selecting the suitable parameters, the four-parameter exponential-type potential [66] can be transformed into the Hulthén potential [60] and the Rosen–Morse potential [67]. A systematic approach was suggested to investigate bound state solutions of exponential-type potentials using the Greene and Aldrich approximation to the centrifugal term [68]. García-Martínez et al. [69] introduced an exactly solvable multi-parameter exponential-type potential by using the canonical transformation technique. This potential has received a lot of interest. Both the Kratzer and Coulomb potentials may be represented as special cases of a general multi-parameter exponential potential using a limiting process [70]. Furthermore, the Mie-type potential can be solved as a particular case of a family of multi-parameter exponential-type potentials in $D$-dimensions without any approximation to the centrifugal term [71].

The multi-parameter exponential-type potential is given in the form [69, 72]

$$u(\beta) = A\frac{e^{-\frac{\beta}{\delta}}}{1-e^{-\frac{\beta}{\delta}}} + B\frac{e^{-\frac{\beta}{\delta}}}{\left(1-e^{-\frac{\beta}{\delta}}\right)^2} + C\frac{e^{-\frac{2\beta}{\delta}}}{\left(1-e^{-\frac{\beta}{\delta}}\right)^2}, \tag{84}$$

where $A$, $B$, $C$, and $\delta$ are real-valued parameters. This potential can be turned into Kratzer and Coulomb potentials by making a Taylor series expansion in the variable $\beta/\delta$, appropriately chosen $A, B, C$ such that $A = a\delta^{-1}, B = a\delta^{-2}, C = c\delta^{-2}$ and performing the limit of the potential when $\delta \to \infty$.

For the multi-parameter exponential-type potential, (18) can be rewritten as

$$\left[\frac{d^2}{d\beta^2} + \epsilon_E - \frac{\nu+2}{\beta^2} - A\frac{e^{-\frac{\beta}{\delta}}}{1-e^{-\frac{\beta}{\delta}}} - B\frac{e^{-\frac{\beta}{\delta}}}{\left(1-e^{-\frac{\beta}{\delta}}\right)^2} - C\frac{e^{-\frac{2\beta}{\delta}}}{\left(1-e^{-\frac{\beta}{\delta}}\right)^2}\right]\bar{\xi}_{n_w,L}(\beta) = 0. \tag{85}$$

Applying the approximation

$$\frac{1}{\beta^2} \approx a^2\left(\mathfrak{d}_0 + \frac{\mathfrak{d}_1}{e^{\frac{\beta}{\delta}}-1} + \frac{\mathfrak{d}_2}{\left(e^{\frac{\beta}{\delta}}-1\right)^2}\right), \tag{86}$$

with $a = \frac{1}{\delta}$, $\mathfrak{d}_0 = \frac{1}{12}$, $\mathfrak{d}_1 = 1$, $\mathfrak{d}_2 = 1$, and then using the definition of the new variable $z = e^{-\frac{\beta}{\delta}}$, so (85) becomes

$$\left[\frac{d^2}{dz^2} + \frac{(1-z)}{(z-z^2)}\frac{d}{dz} + \frac{-\omega_1 z^2 + \omega_2 z - \omega_3}{(z-z^2)^2}\right]\bar{\xi}_{n_w,L}(z) = 0, \tag{87}$$

where

$$\omega_1 = \frac{(\nu+2)}{12} - \delta^2(\epsilon_E + A - C), \tag{88}$$

$$\omega_2 = -\frac{5}{6}(\nu+2) - \delta^2(2\epsilon_E + A + B), \tag{89}$$

$$\omega_3 = \frac{(\nu+2)}{12} - \delta^2\epsilon_E. \tag{90}$$

Rewriting (87) in the conformable fractional form:

$$\mathfrak{D}^\alpha\mathfrak{D}^\alpha\bar{\xi}_{n_w,L}(z) + \frac{(1-z^\alpha)}{(z^\alpha-z^{2\alpha})}\mathfrak{D}^\alpha\bar{\xi}_{n_w,L}(z) + \frac{-\omega_1 z^{2\alpha} + \omega_2 z^\alpha - \omega_3}{(z^\alpha-z^{2\alpha})^2}\bar{\xi}_{n_w,L}(z) = 0. \tag{91}$$

Comparing (91) with (2), the relevant functions $\tilde{\tau}(z)$, $\sigma(z)$, and $\tilde{\sigma}(z)$ are

$$\tilde{\tau}(z) = 1 - z^\alpha, \quad \sigma(z) = z^\alpha - z^{2\alpha}, \quad \tilde{\sigma}(z) = -\omega_1 z^{2\alpha} + \omega_2 z^\alpha - \omega_3. \tag{92}$$

In the CFNU method, the functions $\pi(z)$ are given, according to the same strategy applied in the above cases, as

$$\pi(z) = \frac{(\alpha-1)+(1-2\alpha)z^\alpha}{2} \pm \sqrt{(\mu_1 - \delta^2\epsilon_E - k)z^{2\alpha} - (\mu_2 - 2\delta^2\epsilon_E - k)z^\alpha + (\mu_3 - \delta^2\epsilon_E)}, \tag{93}$$

where $\mu_1 = \frac{(1-2\alpha)^2}{4} + \frac{(\nu+2)}{12} - \delta^2(A-C)$, $\mu_2 = -\frac{(\alpha-1)(1-2\alpha)}{2} - \frac{5}{6}(\nu+2) - \delta^2(A+B)$, and $\mu_3 = \frac{(\alpha-1)^2}{4} + \frac{(\nu+2)}{12}$.

Setting the discriminant of the equation under the square root sign of (93) to be zero, the double roots of $k$ are

$$k_\pm = \mu_2 - 2\mu_3 \pm 2\gamma\Theta, \tag{94}$$

where $\Theta = \sqrt{\mu_3 - \delta^2\epsilon_E}$, and $\gamma = \sqrt{\mu_1 - \mu_2 + \mu_3}$. The possible forms of $\pi(z)$, depending on the two roots of $k_\pm$, are expressed as follows

$$\pi(z) = \frac{(\alpha-1)+(1-2\alpha)z^\alpha}{2} \pm \begin{cases} (\Theta-\gamma)z^\alpha - \Theta, & \text{for } k_+ = \mu_2 - 2\mu_3 + 2\gamma\Theta, \\ (\Theta+\gamma)z^\alpha - \Theta, & \text{for } k_- = \mu_2 - 2\mu_3 - 2\gamma\Theta. \end{cases} \tag{95}$$

The most suitable form of the function $\pi(z)$ must be chosen in order to get a physically acceptable solution; in this case

$$\pi(z) = \frac{(\alpha-1)+(1-2\alpha)z^\alpha}{2} - (\Theta+\gamma)z^\alpha + \Theta, \quad \text{for } k_- = \mu_2 - 2\mu_3 - 2\gamma\Theta, \tag{96}$$

such that the function $\tau(z)$ and its conformable fractional derivative are formulated as

$$\tau(z) = -2(\alpha+\Theta+\gamma)z^\alpha + (\alpha+2\Theta), \tag{97}$$
$$\mathfrak{D}^\alpha\tau(z) = -2\alpha(\alpha+\Theta+\gamma) < 0.$$

Hereafter, the expressions of $\lambda$ and $\lambda_n$ are attained from (10) and (11), respectively, as

$$\lambda = (\mu_2 - 2\mu_3) - 2\gamma\Theta - \alpha\left(\alpha - \frac{1}{2}\right) - \alpha(\Theta + \gamma), \tag{98}$$

$$\lambda_n = n(n-1)\alpha^2 + 2n\alpha(\alpha + \Theta + \gamma). \tag{99}$$

Then the energy eigenvalue solution is achieved by equating equations (98) and (99):

$$\epsilon_E = \frac{(\alpha - 1)^2}{4\delta^2} + \frac{(\nu + 2)}{12\delta^2} - \frac{1}{\delta^2}\left[\frac{n(n+1)\alpha^2 + \alpha\left(\alpha - \frac{1}{2}\right) + (1 + 2n)\alpha\gamma + 2\mu_3 - \mu_2}{(1 + 2n)\alpha + 2\gamma}\right]^2, \tag{100}$$

with $\gamma = \sqrt{\frac{\alpha^2}{4} + (\nu + 2) + \delta^2(B + C)}$. The evolution of the Z(5)-CFE energy levels of the ground, $\beta$, and $\gamma$ bands as a function of the potential parameters and angular momenta are shown in Figs. 5 and 6, respectively.

For the special case $\alpha = 1$, the analytical expression of the energy spectrum reads

$$\epsilon_E = \frac{(\nu + 2)}{12\delta^2} - \frac{1}{\delta^2}\left[\frac{n(n+1) + \frac{1}{2} + (1 + 2n)\sqrt{\frac{1}{4} + (\nu + 2) + \delta^2(B + C)} + (\nu + 2) + \delta^2(A + B)}{(1 + 2n) + 2\sqrt{\frac{1}{4} + (\nu + 2) + \delta^2(B + C)}}\right]^2, \tag{101}$$

which is entirely compatible with (18) of Ref. [72].

Now, the corresponding wave function $\bar{\xi}_{n,n_w,L}(\beta)$, can be achieved by considering

$$\bar{\xi}_{n,n_w,L}(\beta) = \phi(\beta)\chi_{n,\alpha}(\beta), \tag{102}$$

and inserting the expressions for $\sigma(\beta)$ and $\pi(\beta)$ into (5), one obtains

$$\frac{\mathcal{D}^\alpha \phi(\beta)}{\phi(\beta)} = \frac{\pi(\beta)}{\sigma(\beta)} = \frac{\frac{(\alpha-1)}{2} + \Theta - \left(\left(\alpha - \frac{1}{2}\right) + \Theta + \gamma\right)e^{-\frac{\alpha\beta}{\delta}}}{e^{-\frac{\alpha\beta}{\delta}}\left(1 - e^{-\frac{\alpha\beta}{\delta}}\right)}. \tag{103}$$

The solution of (103) can be derived using the same procedure as in the preceding cases, so $\phi(\beta)$ can be written in the form

$$\phi(\beta) = e^{-\frac{1}{\delta}\left(\frac{(\alpha-1)}{2} + \Theta\right)\beta}\left(1 - e^{-\frac{\alpha\beta}{\delta}}\right)^{\left(\frac{1}{2} + \frac{\gamma}{\alpha}\right)}. \tag{104}$$

On the other hand, considering (8), the weight function $\rho(\beta)$ can be expressed as:

$$\rho(\beta) = e^{-\frac{2\Theta}{\delta}\beta}\left(1 - e^{-\frac{\alpha\beta}{\delta}}\right)^{\frac{2\gamma}{\alpha}}. \tag{105}$$

On substituting $\rho(\beta)$ by its formula into (7), we get the function $\chi_{n,\alpha}(\beta)$ as follows

$$\chi_{n,\alpha}(\beta) = \mathcal{B}_{n,n_w,L} e^{\frac{2\Theta}{\delta}\beta}\left(1 - e^{-\frac{\alpha\beta}{\delta}}\right)^{-\frac{2\gamma}{\alpha}}(\mathcal{D}^\alpha)^{(n)}\left(e^{-\frac{1}{\delta}(n\alpha + 2\Theta)\beta}\left(1 - e^{-\frac{\alpha\beta}{\delta}}\right)^{n + \frac{2\gamma}{\alpha}}\right), \quad n = 0,1,2,\ldots, \tag{106}$$

in which $\mathcal{B}_{n,n_w,L}$ are the normalization constants. Finally, the total eigenfunction $\bar{\xi}_{n,n_w,L}(\beta)$ can therefore be established by using (104) and (106) together with the relation (102), as

$$\bar{\xi}_{n,n_w,L}(\beta) = \mathcal{B}_{n,n_w,L}\beta^{-2} e^{-\frac{1}{\delta}\left(\frac{(\alpha-1)}{2} - \Theta\right)\beta}\left(1 - e^{-\frac{\alpha\beta}{\delta}}\right)^{\left(\frac{1}{2} - \frac{\gamma}{\alpha}\right)}(\mathcal{D}^\alpha)^{(n)}\left(e^{-\frac{1}{\delta}(n\alpha + 2\Theta)\beta}\left(1 - e^{-\frac{\alpha\beta}{\delta}}\right)^{n + \frac{2\gamma}{\alpha}}\right), \quad n = 0,1,2,\ldots. \tag{107}$$

For $n = 0$, the analytical expression of the eigenfunction can be written in an explicit form as

$$\bar{\xi}_{0,n_w,L}(\beta) = \mathcal{B}_{0,n_w,L}\beta^{-2} e^{-\frac{1}{\delta}\left(\frac{(\alpha-1)}{2} + \Theta\right)\beta}\left(1 - e^{-\frac{\alpha\beta}{\delta}}\right)^{\left(\frac{1}{2} + \frac{\gamma}{\alpha}\right)}, \tag{108}$$

where the expression of the normalization constant is defined by

$$\mathcal{B}_{0,n_w,L} = \frac{1}{\sqrt{\int_0^\infty e^{-\frac{1}{\delta}((\alpha-1) + 2\Theta)\beta}\left(1 - e^{-\frac{\alpha\beta}{\delta}}\right)^{\left(1 + \frac{2\gamma}{\alpha}\right)} d\beta}}. \tag{109}$$

Table 1
The values of dimensionless parameters of the Z(5)-CFM, Z(5)-CFTH, and Z(5)-CFE models.

| Isotopes | Z(5)-CFM | | | Z(5)-CFTH | | | | Z(5)-CFE | | | | |
|---|---|---|---|---|---|---|---|---|---|---|---|---|
| | $\alpha$ | $\beta_e$ | $a$ | $\alpha$ | $b_h$ | $c_h$ | $\beta_e$ | $\alpha$ | $A$ | $B$ | $C$ | $\delta$ |
| $^{114}$Pd | 0.90 | 4.70 | 0.41 | 0.80 | 0.30 | 0.10 | 5.00 | 1.00 | 12.00 | 0.70 | 0.061 | 10.0 |
| $^{116}$Pd | 1.00 | 4.20 | 0.30 | 0.70 | 0.20 | 0.10 | 5.10 | 1.00 | 20.00 | 0.40 | 0.050 | 13.0 |
| $^{126}$Xe | 1.00 | 4.10 | 0.40 | 1.00 | 0.20 | 0.30 | 4.50 | 0.70 | 28.00 | 0.90 | 0.030 | 8.00 |
| $^{128}$Xe | 1.00 | 3.80 | 0.50 | 1.00 | 0.02 | 0.89 | 5.70 | 1.00 | 28.00 | 0.30 | 0.070 | 10.0 |
| $^{192}$Pt | 1.00 | 4.20 | 0.50 | 0.90 | 0.20 | 0.40 | 5.00 | 0.90 | 28.00 | 0.60 | 0.020 | 9.00 |
| $^{194}$Pt | 0.95 | 4.20 | 0.50 | 0.90 | 0.20 | 0.40 | 5.10 | 0.90 | 7.000 | 0.60 | 0.020 | 9.00 |

## 7. Numerical results

To test the Z(5)-CFM, Z(5)-CFTH, and Z(5)-CFE models, the $^{114,116}$Pd, $^{126,128}$Xe, and $^{192,194}$Pt isotopes have been chosen because they are excellent candidates for triaxial nuclei. The neutron-rich Pd nuclei are of particular interest since a prolate-to-oblate QSPT was predicted using the mean field studies [73]. In addition to the empirical investigations, much effort was put into describing the characteristics of the Pd nuclei within the framework of interacting boson model [74]. In $^{112-116}$Pd, a transition from triaxial prolate to triaxial oblate was found. In $^{112-116}$Pd, doublet bands were seen but exhibited varying softness to triaxial shapes, with wobbling motion seen only in $^{114}$Pd [75]. For $^{192-196}$Pt, which is assumed to be around the prolate to oblate critical point, the predicted energy spectra and $B(E2)$ transition rates using the Z(5) model were in excellent agreement with experimental results [7]. Clear evidence for triaxiality in the $^{120-130}$Xe nuclei near A=130 was presented [76].

Firstly, it is essential to analyze the behavior of the energy spectra and the $B(E2)$ transition rates of the Z(5)-CFM, Z(5)-CFTH, and Z(5)-CFE models as a function of the fractional-order parameter $\alpha$ and the potential parameters. The ground, $\beta$, and $\gamma$ bands are described by a set of quantum numbers $(n, n_\gamma, n_w)$. The ground band is labeled by $(n, n_\gamma, n_w) \equiv (0,0,0)$ [7]. $(n, n_\gamma, n_w) \equiv (1,0,0)$ refers to the $\beta$ band. The even $L$ levels of the $\gamma$ band is characterized by $(n, n_\gamma, n_w) \equiv (0,0,2)$ while the $\gamma$ band with odd $L$ levels is labeled by $(n, n_\gamma, n_w) \equiv (0,0,1)$. The states are denoted by the symbol $L^+_{n,n_w}$. All levels are measured from the state $0^+_{0,0}$. The energy spectrum is normalized to the $2^+_{0,0}$ state as follows,

$$\mathcal{E}^\alpha_{L^+_{n,n_w}} = \frac{\epsilon^\alpha_{L^+_{n,n_w}} - \epsilon^\alpha_{0^+_{0,0}}}{\epsilon^\alpha_{2^+_{0,0}} - \epsilon^\alpha_{0^+_{0,0}}}. \tag{110}$$

However, the $B(E2)$ transition probabilities are normalized to the $B(E2: 2^+_{0,0} \to 0^+_{0,0})$ according to,

$$B(E2) = \frac{B(E2; L_i, \varrho_i \to L_f, \varrho_f)}{B(E2; 2^+_{0,0} \to 0^+_{0,0})}. \tag{111}$$

The root mean square deviation $\sigma$ is utilized to evaluate the quality measure. It is given by

$$\sigma = \sqrt{\frac{\Sigma_i \left(\mathcal{E}^{\exp}_{L^+_{n,n_w}} - \mathcal{E}^\alpha_{L^+_{n,n_w}}\right)^2}{n_l - 1}}, \tag{112}$$

where $n_l$ denotes the number of states in the fitting process, however the theoretical energies, and the corresponding experimental energies of the $L^+_{n,n_w}$ states are indicated by $\mathcal{E}^\alpha_{L^+_{n,n_w}}$ and $\mathcal{E}^{\exp}_{L^+_{n,n_w}}$, respectively.

For $\alpha = 1$ (classical case), Figs. 1(a) and 1(c) represent the evolution of the Z(5)-CFM energy spectra of the ground, $\beta$, and $\gamma$ bands as a function of the parameters $\beta_e \in [0.1,5]$, and $a \in [0.1,0.5]$, respectively. The energy levels are normalized based on (110). In the fractional case, for $\alpha = 0.5$, the development of the low-lying spectra of the Z(5)-CFM model is depicted in Figs. 1(b), and 1(d). By increasing the parameters $\beta_e$ or $a$, it can be seen that the states from the $\beta$ band intersect with the states from the ground and $\gamma$ bands at some points (the levels change their order with changing the parameters $\beta_e$ or $a$). In Fig. 2 the energy levels and $B(E2; L_f + 2 \to L_f)$ rates of the ground band of the Z(5)-CFM model are compared with the Z(5) model predictions for different values of the parameters $\beta_e$ and $a$. It is observed that the energy increases monotonically with increasing $L$. For small values of the angular momentum $L$, the energy levels and $B(E2; L_f + 2 \to L_f)$ rates are close to the predictions of the Z(5) model. However, for large values of $L$, there is a significant deviation from the results of the Z(5) model. Similarly, the evolution of the low-lying energy spectra, the ground band levels, and $B(E2; L_f + 2 \to L_f)$ rates are presented in Figs. 3 and 4 for the Z(5)-CFTH model and in Figs. 5 and 6 for the Z(5)-CFE model. These curves present a behavior similar to those illustrated in Figs. 1 and 2.

Table 1 represents the values of the parameters of the Z(5)-CFM, Z(5)-CFTH, and Z(5)-CFE models which are used to fit the experimental data. Tables 2, 3, and 4 show the experimental results and the theoretical energy levels using the Z(5)-CFM (46),

Z(5)-CFTH (76), Z(5)-CFE (100), Z(5)-CFK ((51) in Ref. [31]), and Z(5) [7] models for the $^{126,128}$Xe, $^{192,194}$Pt, and $^{114,116}$Pd isotopes, respectively. Furthermore, Tables 5, 6, and 7 represent the experimental findings and the theoretical predictions of the $B(E2)$ transition rates of the Z(5)-CFM, Z(5)-CFTH, Z(5)-CFE, Z(5)-CFK, and Z(5) models for the $^{126,128}$Xe, $^{192,194}$Pt, and $^{114,116}$Pd isotopes, respectively. From these Tables, we can see that the values of $\sigma$ are large with the Z(5)-CFM model.

Obviously, from Fig. 1, for the Z(5)-CFM energy spectra, the values of energy levels, for $\alpha = 0.5$, are lower than the corresponding values in the classical case ($\alpha = 1$). This result is very clear for the $\beta$ band. Moreover, this fact is correct in the case of Z(5)-CFTH and Z(5)-CFE models. So, to avoid repetition, we represent the evolution of the Z(5)-CFTH and Z(5)-CFE energy levels of the ground, $\beta$, and $\gamma$ bands against the potentials parameters in the classical case only, $\alpha = 1$. The order of the fractional derivative is a continuous parameter in the CFBH. As a result, the CFBH provides numerous models that are similar and very close to the original one, increasing the chances that a model will fall on the experimental results with a high level of accuracy, mainly if $\alpha$ is close to 1.

In nuclear structure theory, the investigation of nuclei at the CPSs of the QSPT is extremely hard since the structure at these points varies considerably. The Z(5) CPS was introduced to analyze the nuclei at the CPS of the QSPT when it transitioned from the triaxial vibrator to the rigid triaxial rotator. In the new models (Z(5)-CFM, Z(5)-CFTH, and Z(5)-CFE), the infinite potential used in Z(5) CPS is replaced by Morse, Tietz-Hua, and multi-parameter exponential-type potentials, respectively. In addition to providing analytical solutions for the $B(E2)$ transition rates and the energy spectra that can be readily compared to empirical observations (by varying the parameters of the potentials), for the diversity of triaxial nuclei, the suggested technique offers an intriguing by-product. The Z(5) solution is a particular instance of the Z(5)-CFTH, and Z(5)-CFE models that are generated for certain parameter values. Figs. 4, and 6 represent this result, especially for $L \leq 12$. Consequently, these models may cover a vast region of triaxial nuclei and, in particular, determine the Z(5) CPSs.

The results of the Z(5)-CFE and Z(5)-CFK models are very close and show good fitting with experimental data (the values of $\sigma$ are the smallest). This result is expected because the Kratzer potential may be represented as a particular case of a general multi-parameter exponential potential using a limiting process [70]. On the other hand, Z(5)-CFK and traditional Z(5) model outcomes are comparably similar. This is because of the geometrical shape of the Kratzer potential. For appropriate values of its parameters, the Kratzer potential may take on the form of a deep well, analogous to the infinite square-well potential of the Z(5) model, which is frequently used to describe the $^{192-196}$Pt isotopes.

We can see that the predictions of the normalized $B(E2)$ transition rates as well as the normalized energy spectra of the ground, $\beta$, and $\gamma$ bands provided by the CFBH (with $\alpha \in [0.7,1]$) are more accurate than the results produced from the classical collective BH (depending on the values of $\sigma$).

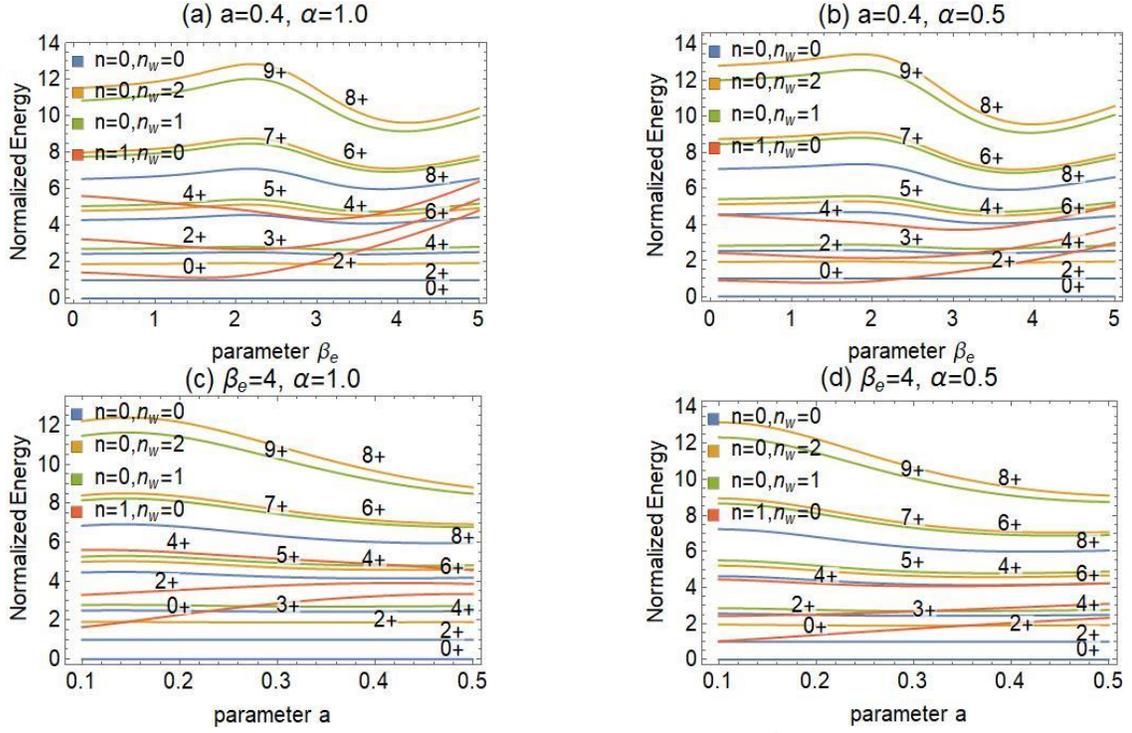

Fig. 1. (a) and (b) represent the evolution of the Z(5)-CFM energy levels of the ground, $\beta$, and $\gamma$ bands against the parameter $\beta_e$, $\beta_e \in [0.1, 5]$, $\alpha = 1$ (classical case), and $\alpha = 0.5$ (fractional case), respectively. The bands are described by $(n, n_w)$. The inset graphic legends show the bands. (c) and (d) represent the development of the Z(5)-CFM energy levels against the parameter $a$, $a \in [0.1, 0.5]$, $\alpha = 1$ (classical case), and $\alpha = 0.5$ (fractional case), respectively. The energy levels are normalized based on (110).

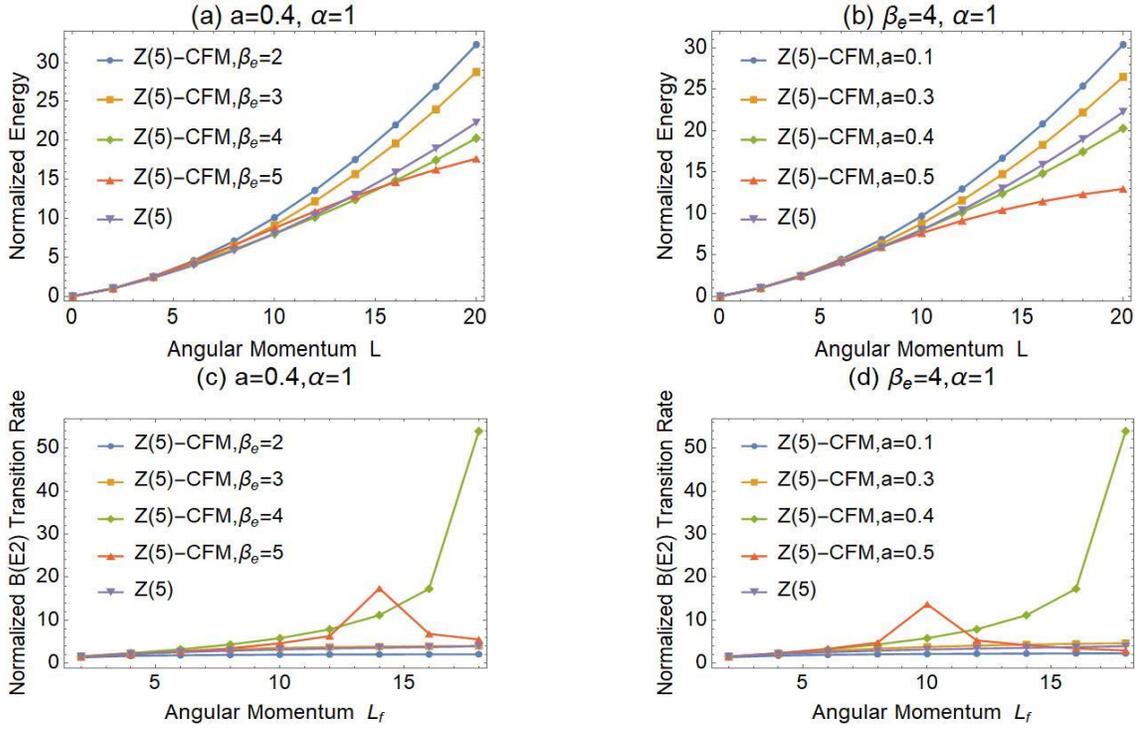

Fig. 2. (a) and (b) represent the behavior of the Z(5)-CFM energy levels against the angular momentum $L$. The energy spectra are normalized based on (110). However, (c) and (d) are the evolution of $B(E2; L_f + 2 \rightarrow L_f)$ rates for the ground bands of the Z(5)-CFM model against the angular momentum $L_f$. The $B(E2)$ rates are normalized based on (111). The energy levels and $B(E2; L_f + 2 \rightarrow L_f)$ rates are compared with the Z(5) model predictions.

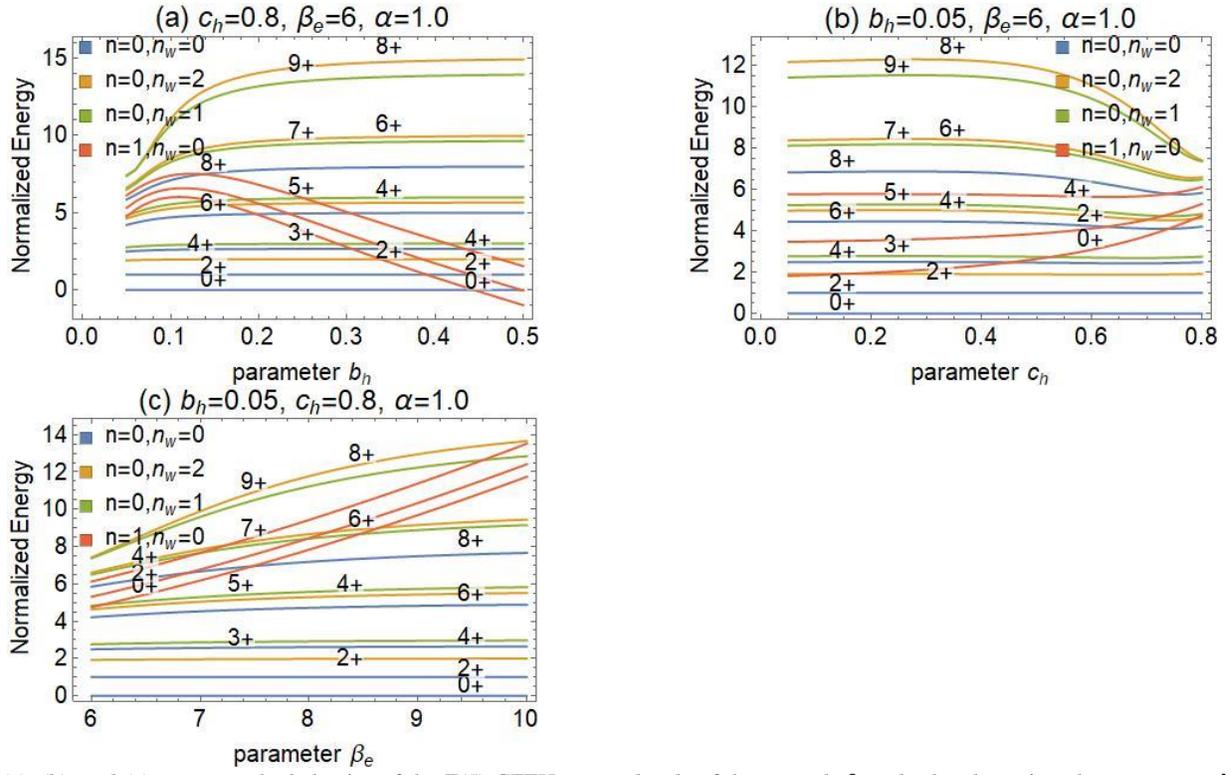

Fig. 3. (a), (b), and (c) represent the behavior of the Z(5)-CFTH energy levels of the ground, $\beta$, and $\gamma$ bands against the parameters $b_h \in [0.05, 0.5]$, $c_h \in [0.02, 0.8]$, and $\beta_e \in [6, 10]$, respectively. The bands are described by $(n, n_w)$. The inset graphic legends show the bands.

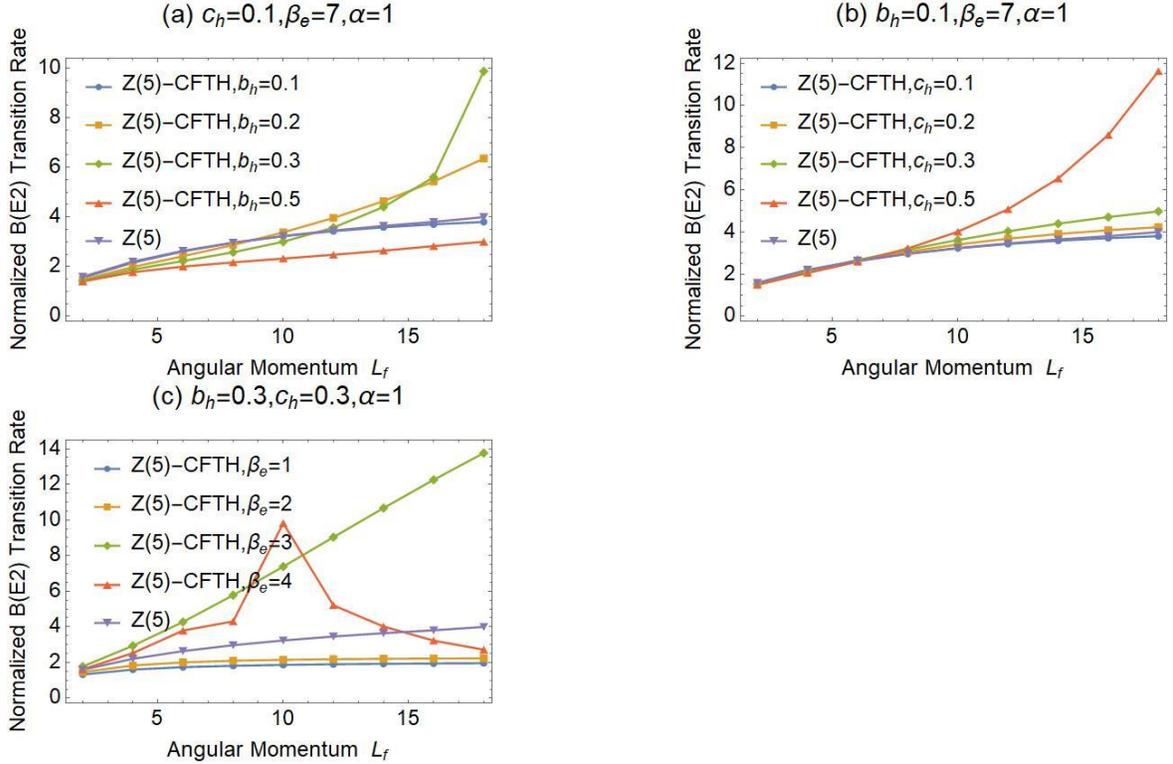

Fig. 4. (a), (b), and (c) represent the evolution of $B(E2; L_f + 2 \to L_f)$ rates for the ground bands of the Z(5)-CFTH model as a function of the parameters $\alpha$, $c_h$, $b_h$ and $\beta_e$, ((a) $c_h = 0.1$, $\beta_e = 7$, $\alpha = 1$; $b_h = 0.1, 0.2, 0.3, 0.5$, (b) $b_h = 0.1$, $\beta_e = 7$, $\alpha = 1$; $c_h = 0.1, 0.2, 0.3, 0.5$, and $b_h = 0.3$, $c_h = 0.3$, $\alpha = 1$; $\beta_e = 1, 2, 3, 4$, for (c)) against the values of the angular momentum $L_f$. The $B(E2)$ rates are normalized based on (111). The $B(E2; L_f + 2 \to L_f)$ rates are compared with the Z(5) model predictions.

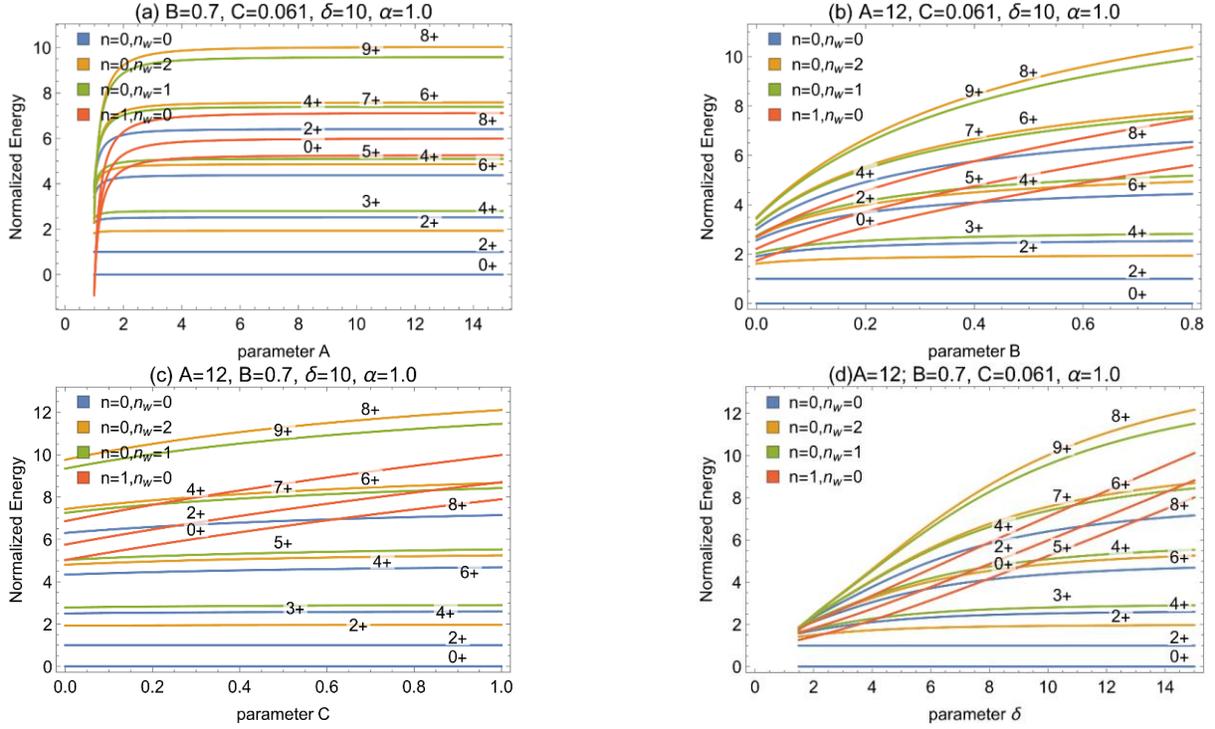

Fig. 5. (a), (b), (c), and (d) are the plots of the low-lying energy levels of the ground, $\beta$, and $\gamma$ bands of the Z(5)-CFE potential against the parameters $A \in [1,15]$, $B \in [0,0.8]$, $C \in [0,1.0]$, and $\delta \in [2,15]$, respectively. The bands are described by $(n, n_w)$. The inset graphic legends show the bands.

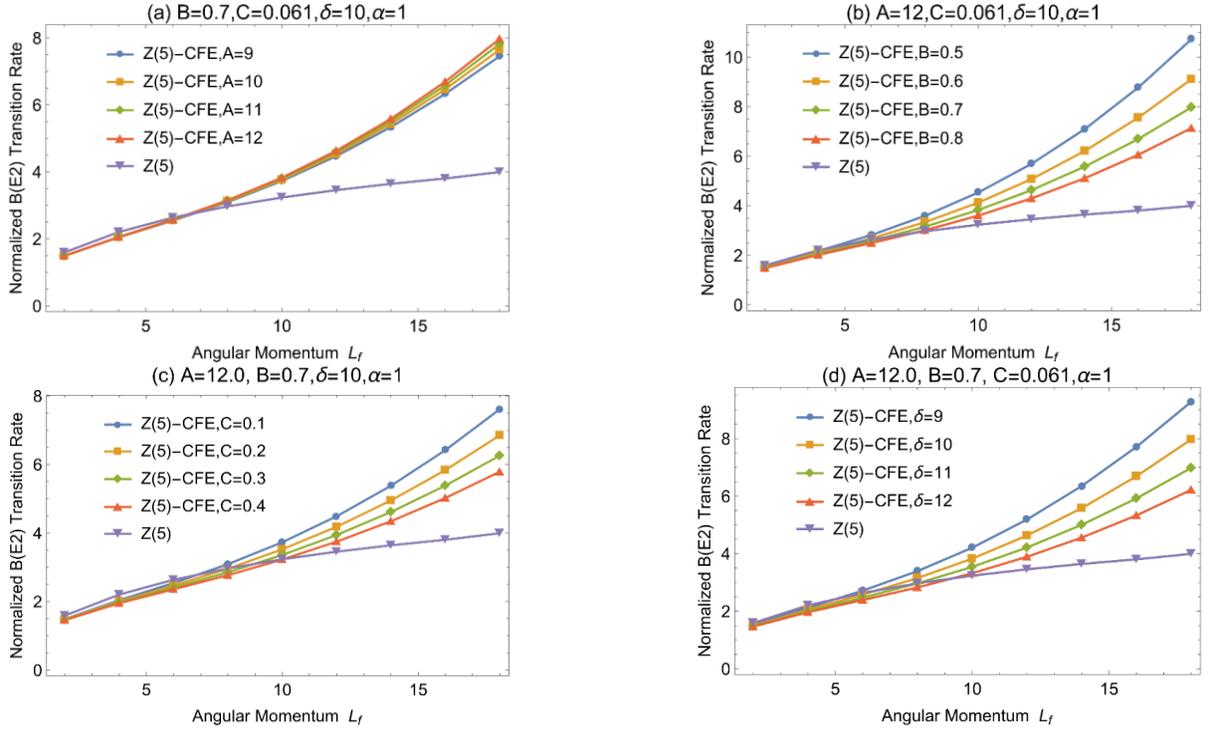

Fig. 6. (a), (b), (c), and (d) represent the evolution of $B(E2; L_f + 2 \to L_f)$ rates for the ground bands of the Z(5)-CFE model as a function of the parameters $A$, $B$, $C$, $\delta$ and $\alpha$, ((a) $B = 0.7$, $C = 0.061$, $\delta = 10$, $\alpha = 1$; $A = 9, 10, 11, 12$, (b) $A = 12$, $C = 0.061$, $\delta = 10$, $\alpha = 1$; $B = 0.5, 0.6, 0.7, 0.8$, (c) $A = 12$, $B = 0.7$, $\delta = 10$, $\alpha = 1$; $C = 0.1, 0.2, 0.3, 0.4$ and $A = 12$, $B = 0.7$, $C = 0.061$, $\alpha = 1$; $\delta = 9, 10, 11, 12$ for (d)) against the values of the angular momentum $L_f$. The $B(E2)$ rates are normalized based on (111). The $B(E2; L_f + 2 \to L_f)$ rates are compared with the Z(5) model predictions.

Table 2
Comparison of theoretical energy levels predictions of the Z(5)-CFM, Z(5)-CFTH, Z(5)-CFE, and Z(5)-CFK models, given by (46), (76), (100), and ((51) in Ref. [31]), respectively, with the predictions of Z(5) model [7] and the experimental results for $^{126}$Xe [77] and $^{128}$Xe [78]. The energy levels are normalized based on (110). The values of the parameters are shown in Table 1.

| $L^+_{n,n_w}$ | $^{126}$Xe | | | | | $^{128}$Xe | | | | | Z(5) |
| --- | --- | --- | --- | --- | --- | --- | --- | --- | --- | --- | --- |
| | Exp. | Z(5)-CFM | Z(5)-CFTH | Z(5)-CFE | Z(5)-CFK | Exp. | Z(5)-CFM | Z(5)-CFTH | Z(5)-CFE | Z(5)-CFK | |
| $2^+_{0,0}$ | 1.000 | 1.0000 | 1.0000 | 1.0000 | 1.0000 | 1.000 | 1.0000 | 1.0000 | 1.0000 | 1.0000 | 1.000 |
| $4^+_{0,0}$ | 2.424 | 2.4522 | 2.4210 | 2.4806 | 2.4798 | 2.333 | 2.4422 | 2.3929 | 2.3981 | 2.3982 | 2.350 |
| $6^+_{0,0}$ | 4.207 | 4.1730 | 4.0767 | 4.2376 | 4.2347 | 3.922 | 4.1027 | 3.9324 | 3.9417 | 3.9419 | 3.984 |
| $8^+_{0,0}$ | 6.267 | 6.0452 | 5.8673 | 6.0848 | 6.0785 | 5.674 | 5.8044 | 5.4478 | 5.4449 | 5.4454 | 5.877 |
| $10^+_{0,0}$ | 8.645 | 8.0197 | 7.7633 | 7.8860 | 7.8753 | 7.597 | 7.4413 | 6.8542 | 6.8059 | 6.8068 | 8.019 |
| $12^+_{0,0}$ | 10.99 | 10.091 | 9.7713 | 9.5584 | 9.5431 | | | | | | 10.40 |
| $2^+_{0,2}$ | 2.264 | 1.8987 | 1.8828 | 1.9139 | 1.9135 | 2.189 | 1.8952 | 1.8711 | 1.8740 | 1.8741 | 1.837 |
| $4^+_{0,2}$ | 3.830 | 4.6162 | 4.5008 | 4.6836 | 4.6800 | 3.620 | 4.5170 | 4.3068 | 4.3154 | 4.3156 | 4.420 |
| $6^+_{0,2}$ | 5.698 | 7.1547 | 6.9309 | 7.1196 | 7.1109 | 5.150 | 6.7458 | 6.2622 | 6.2386 | 6.2393 | 7.063 |
| $8^+_{0,2}$ | 7.878 | 9.6314 | 9.3238 | 9.2079 | 9.1935 | | | | | | 9.864 |
| $3^+_{0,1}$ | 3.391 | 2.7172 | 2.6774 | 2.7523 | 2.7513 | 3.228 | 2.7023 | 2.6384 | 2.6446 | 2.6447 | 2.597 |
| $5^+_{0,1}$ | 4.898 | 4.8313 | 4.7065 | 4.8985 | 4.8945 | 4.508 | 4.7157 | 4.4851 | 4.4929 | 4.4932 | 4.634 |
| $7^+_{0,1}$ | 6.848 | 6.9757 | 6.7591 | 6.9564 | 6.9481 | 6.165 | 6.5976 | 6.1351 | 6.1156 | 6.1163 | 6.869 |
| $9^+_{0,1}$ | 9.059 | 9.1617 | 8.8677 | 8.8375 | 8.8242 | | | | | | 9.318 |
| $0^+_{1,0}$ | 3.381 | 3.3749 | 3.3947 | 3.3936 | 3.3854 | 3.575 | 3.0551 | 3.5692 | 3.6495 | 3.6501 | 3.913 |
| $2^+_{1,0}$ | 4.319 | 4.0386 | 4.1361 | 4.1774 | 4.1691 | 4.515 | 3.5699 | 4.2449 | 4.3150 | 4.3156 | 5.697 |
| $4^+_{1,0}$ | 5.255 | 5.0021 | 5.2192 | 5.3458 | 5.3367 | | | | | | 7.962 |
| $\sigma$ | | 0.7001 | 0.6751 | 0.6994 | 0.69 | | 0.6579 | 0.4864 | 0.4864 | 0.48 | |

Table 3
Same as Table 2 but for the $^{192}$Pt [79] and $^{194}$Pt [80].

| $L^+_{n,n_w}$ | $^{192}$Pt | | | | | $^{194}$Pt | | | | | Z(5) |
| --- | --- | --- | --- | --- | --- | --- | --- | --- | --- | --- | --- |
| | Exp. | Z(5)-CFM | Z(5)-CFTH | Z(5)-CFE | Z(5)-CFK | Exp. | Z(5)-CFM | Z(5)-CFTH | Z(5)-CFE | Z(5)-CFK | |
| $2^+_{0,0}$ | 1.000 | 1.0000 | 1.0000 | 1.0000 | 1.0000 | 1.000 | 1.0000 | 1.0000 | 1.0000 | 1.0000 | 1.000 |
| $4^+_{0,0}$ | 2.479 | 2.4932 | 2.4790 | 2.4546 | 2.4541 | 2.470 | 2.4943 | 2.4899 | 2.4531 | 2.4506 | 2.350 |
| $6^+_{0,0}$ | 4.314 | 4.2734 | 4.2023 | 4.1419 | 4.1401 | 4.299 | 4.2781 | 4.2430 | 4.1356 | 4.1274 | 3.984 |
| $8^+_{0,0}$ | 6.377 | 6.1321 | 5.9153 | 5.8724 | 5.8688 | 6.392 | 6.1439 | 6.0049 | 5.8569 | 5.8408 | 5.877 |
| $10^+_{0,0}$ | 8.624 | 7.8941 | 7.3824 | 7.5189 | 7.5129 | 8.671 | 7.9175 | 7.5299 | 7.4887 | 7.4651 | 8.019 |
| $12^+_{0,0}$ | | | | | | | | | | | 10.40 |
| $2^+_{0,2}$ | 1.935 | 1.9202 | 1.9141 | 1.9014 | 1.9012 | 1.894 | 1.9208 | 1.9193 | 1.9007 | 1.8995 | 1.837 |
| $4^+_{0,2}$ | 3.795 | 4.7246 | 4.6284 | 4.5637 | 4.5616 | 3.743 | 4.7307 | 4.6794 | 4.5557 | 4.5456 | 4.420 |
| $6^+_{0,2}$ | 5.905 | 7.1537 | 6.7919 | 6.8231 | 6.8181 | 5.863 | 7.1715 | 6.9144 | 6.8000 | 6.7793 | 7.063 |
| $8^+_{0,2}$ | 8.186 | 9.1218 | 8.2326 | 8.7026 | 8.6947 | 8.186 | 9.1576 | 8.4194 | 8.6569 | 8.6295 | 9.864 |
| $3^+_{0,1}$ | 2.910 | 2.7683 | 2.7487 | 2.7182 | 2.7176 | 2.809 | 2.7698 | 2.7630 | 2.7162 | 2.7130 | 2.597 |
| $5^+_{0,1}$ | 4.682 | 4.9416 | 4.8313 | 4.7660 | 4.7637 | 4.563 | 4.9483 | 4.8877 | 4.7570 | 4.7461 | 4.634 |
| $7^+_{0,1}$ | 6.677 | 6.9940 | 6.6590 | 6.6741 | 6.6693 | | | | | | 6.869 |
| $9^+_{0,1}$ | | | | | | | | | | | 9.318 |
| $0^+_{1,0}$ | 3.776 | 3.6570 | 4.0429 | 3.8855 | 3.8803 | 3.858 | 3.6198 | 4.1825 | 3.8644 | 3.8414 | 3.913 |
| $2^+_{1,0}$ | 4.547 | 4.1713 | 4.6690 | 4.6057 | 4.6004 | 4.603 | 4.1639 | 4.8087 | 4.5812 | 4.5595 | 5.697 |
| $4^+_{1,0}$ | 6.110 | 4.8935 | 5.5792 | 5.6640 | 5.6582 | 5.817 | 4.9346 | 5.7253 | 5.6330 | 5.6121 | 7.962 |
| $\sigma$ | | 0.6390 | 0.5118 | 0.4976 | 0.49 | | 0.6469 | 0.5375 | 0.5227 | 0.52 | |

Table 4
Same as Table 2 but for the $^{114}$Pd [81] and $^{116}$Pd [82].

| $L^+_{n,n_w}$ | $^{114}$Pd | | | | | $^{116}$Pd | | | | | Z(5) |
|---|---|---|---|---|---|---|---|---|---|---|---|
| | Exp. | Z(5)-CFM | Z(5)-CFTH | Z(5)-CFE | Z(5)-CFK | Exp. | Z(5)-CFM | Z(5)-CFTH | Z(5)-CFE | Z(5)-CFK | |
| $2^+_{0,0}$ | 1.000 | 1.0000 | 1.0000 | 1.0000 | 1.0000 | 1.000 | 1.0000 | 1.0000 | 1.0000 | 1.0000 | 1.000 |
| $4^+_{0,0}$ | 2.563 | 2.4456 | 2.4585 | 2.5179 | 2.5179 | 2.518 | 2.4376 | 2.4532 | 2.4537 | 2.4532 | 2.350 |
| $6^+_{0,0}$ | 4.511 | 4.2181 | 4.2453 | 4.3792 | 4.3791 | 4.415 | 4.0974 | 4.1368 | 4.1384 | 4.1368 | 3.984 |
| $8^+_{0,0}$ | 6.662 | 6.2828 | 6.3088 | 6.4104 | 6.4104 | 6.585 | 5.8228 | 5.8621 | 5.8645 | 5.8613 | 5.877 |
| $10^+_{0,0}$ | 8.598 | 8.6463 | 8.6449 | 8.4686 | 8.4687 | 8.164 | 7.5314 | 7.5035 | 7.5050 | 7.4999 | 8.019 |
| $12^+_{0,0}$ | | | | | | | | | | | 10.40 |
| $2^+_{0,2}$ | 2.088 | 1.8921 | 1.8994 | 1.9315 | 1.9315 | 2.146 | 1.8925 | 1.9007 | 1.9010 | 1.9008 | 1.837 |
| $4^+_{0,2}$ | 3.968 | 4.6924 | 4.7213 | 4.8625 | 4.8625 | 3.861 | 4.5144 | 4.5575 | 4.5593 | 4.5574 | 4.420 |
| $6^+_{0,2}$ | | | | | | 5.549 | 6.7976 | 6.8097 | 6.8120 | 6.8077 | 7.063 |
| $8^+_{0,2}$ | | | | | | | | | | | 9.864 |
| $3^+_{0,1}$ | 3.042 | 2.7128 | 2.7284 | 2.8014 | 2.8013 | 3.124 | 2.6968 | 2.7164 | 2.7170 | 2.7164 | 2.597 |
| $5^+_{0,1}$ | 4.903 | 4.9256 | 4.9549 | 5.0969 | 5.0969 | 4.901 | 4.7149 | 4.7592 | 4.7611 | 4.7590 | 4.634 |
| $7^+_{0,1}$ | | | | | | | | | | | 6.869 |
| $9^+_{0,1}$ | | | | | | | | | | | 9.318 |
| $0^+_{1,0}$ | | | | | | | | | | | 3.913 |
| $2^+_{1,0}$ | | | | | | | | | | | 5.697 |
| $4^+_{1,0}$ | | | | | | | | | | | 7.962 |
| $\sigma$ | | 0.3387 | 0.3377 | 0.3570 | 0.36 | | 0.6106 | 0.6111 | 0.6112 | 0.61 | |

Table 5
Comparison of theoretical predictions of the $B(E2)$ transition rates of the Z(5)-CFM, Z(5)-CFTH, Z(5)-CFE, and Z(5)-CFK models with the predictions of Z(5) model [7], and the experimental results for $^{126}$Xe [77] and $^{128}$Xe [78]. The $B(E2)$ transition rates are normalized based on (111).

| $L^{(i)}_{n,n_w}$ | $L^{(f)}_{n,n_w}$ | $^{126}$Xe | | | | | $^{128}$Xe | | | | | Z(5) |
|---|---|---|---|---|---|---|---|---|---|---|---|---|
| | | Exp. | Z(5)-CFM | Z(5)-CFTH | Z(5)-CFE | Z(5)-CFK | Exp. | Z(5)-CFM | Z(5)-CFTH | Z(5)-CFE | Z(5)-CFK | |
| $4_{0,0}$ | $2_{0,0}$ | | 1.5840 | 1.6027 | 1.5198 | 1.6032 | 1.468 | 1.5989 | 1.6096 | 1.5967 | 1.6033 | 1.590 |
| $6_{0,0}$ | $4_{0,0}$ | | 2.3619 | 2.4047 | 2.1544 | 2.4512 | 1.941 | 2.4589 | 2.4594 | 2.4171 | 2.4428 | 2.203 |
| $8_{0,0}$ | $6_{0,0}$ | | 3.2301 | 3.2554 | 2.7944 | 3.4915 | 2.388 | 3.6193 | 3.4631 | 3.3762 | 3.4428 | 2.635 |
| $4_{0,2}$ | $2_{0,2}$ | | 0.7548 | 0.7677 | 0.6907 | 0.7745 | | 0.7564 | 0.7652 | 0.7574 | 0.7596 | 0.736 |
| $6_{0,2}$ | $4_{0,2}$ | | 1.3031 | 1.3107 | 1.0963 | 1.4134 | | 1.3770 | 1.3871 | 1.3517 | 1.3752 | 1.031 |
| $8_{0,2}$ | $6_{0,2}$ | | 2.5985 | 2.4443 | 2.0287 | 3.0215 | | 4.7450 | 2.8491 | 2.7616 | 2.8586 | 1.590 |
| $5_{0,1}$ | $3_{0,1}$ | | 1.3467 | 1.3710 | 1.2110 | 1.4048 | | 1.4018 | 1.4013 | 1.3765 | 1.3907 | 1.235 |
| $7_{0,1}$ | $5_{0,1}$ | | 2.4021 | 2.4001 | 2.0276 | 2.6365 | | 2.7262 | 2.5895 | 2.5200 | 2.5756 | 1.851 |
| $9_{0,1}$ | $7_{0,1}$ | | 3.7224 | 3.5095 | 2.9461 | 4.3358 | | 3.1354 | 4.1032 | 3.9788 | 4.1250 | 2.308 |
| $2_{0,2}$ | $2_{0,0}$ | | 1.6046 | 1.6239 | 1.5501 | 1.6181 | 1.194 | 1.6186 | 1.6319 | 1.6184 | 1.6257 | 1.620 |
| $4_{0,2}$ | $4_{0,0}$ | | 0.3725 | 0.3792 | 0.3375 | 0.3865 | | 0.3843 | 0.3854 | 0.3793 | 0.3826 | 0.348 |
| $6_{0,2}$ | $6_{0,0}$ | | 0.2368 | 0.2407 | 0.1996 | 0.2518 | | 0.2300 | 0.2471 | 0.2415 | 0.2440 | 0.198 |
| $3_{0,1}$ | $4_{0,0}$ | | 1.3032 | 1.3286 | 1.2128 | 1.3396 | | 1.3517 | 1.3598 | 1.3361 | 1.3513 | 1.243 |
| $5_{0,1}$ | $6_{0,0}$ | | 1.1749 | 1.1862 | 1.0373 | 1.2669 | | 1.3369 | 1.2717 | 1.2384 | 1.2662 | 0.972 |
| $7_{0,1}$ | $8_{0,0}$ | | 1.1736 | 1.1395 | 0.9826 | 1.3444 | | 1.6813 | 1.3081 | 1.2687 | 1.3142 | 0.808 |
| $3_{0,1}$ | $2_{0,2}$ | | 2.2301 | 2.2703 | 2.0944 | 2.2787 | | 2.2884 | 2.3069 | 2.2730 | 2.2934 | 2.171 |
| $5_{0,1}$ | $4_{0,2}$ | | 1.6248 | 1.6345 | 1.4269 | 1.7668 | | 1.8935 | 1.7705 | 1.7221 | 1.7648 | 1.313 |
| $7_{0,1}$ | $6_{0,2}$ | | 1.9629 | 1.8643 | 1.6219 | 2.3024 | | 3.4222 | 2.2114 | 2.1440 | 2.2354 | 1.260 |

Table 6
Same as Table 5 but for the $^{192}$Pt [79] and $^{194}$Pt [80].

| | | $^{192}$Pt | | | | | $^{194}$Pt | | | | | |
|---|---|---|---|---|---|---|---|---|---|---|---|---|
| $L^{(i)}_{n,n_w}$ | $L^{(f)}_{n,n_w}$ | Exp. | Z(5)-CFM | Z(5)-CFTH | Z(5)-CFE | Z(5)-CFK | Exp. | Z(5)-CFM | Z(5)-CFTH | Z(5)-CFE | Z(5)-CFK | Z(5) |
| $4_{0,0}$ | $2_{0,0}$ | 1.559 | 1.5467 | 1.5394 | 1.5464 | 1.5763 | 1.728 | 1.5522 | 1.5310 | 1.5273 | 1.5796 | 1.590 |
| $6_{0,0}$ | $4_{0,0}$ | 1.224 | 2.2750 | 2.2429 | 2.2434 | 2.3502 | 1.362 | 2.2939 | 2.2138 | 2.1740 | 2.3616 | 2.203 |
| $8_{0,0}$ | $6_{0,0}$ | | 3.1767 | 3.0642 | 2.9872 | 3.2386 | 1.016 | 3.2190 | 2.9976 | 2.8214 | 3.2647 | 2.635 |
| $4_{0,2}$ | $2_{0,2}$ | | 0.7222 | 0.7071 | 0.7193 | 0.7463 | 0.446 | 0.7258 | 0.7025 | 0.7103 | 0.7485 | 0.736 |
| $6_{0,2}$ | $4_{0,2}$ | | 1.2239 | 1.1808 | 1.1873 | 1.2973 | | 1.2377 | 1.1569 | 1.1249 | 1.3080 | 1.031 |
| $8_{0,2}$ | $6_{0,2}$ | | 2.2413 | 2.2862 | 2.2734 | 2.6235 | | 0.4011 | 2.2170 | 2.0539 | 2.6571 | 1.590 |
| $5_{0,1}$ | $3_{0,1}$ | | 1.2921 | 1.2658 | 1.2700 | 1.3380 | | 1.3031 | 1.2491 | 1.2304 | 1.3449 | 1.235 |
| $7_{0,1}$ | $5_{0,1}$ | | 2.3655 | 2.2583 | 2.1938 | 2.4107 | | 2.3981 | 2.2044 | 2.0564 | 2.4325 | 1.851 |
| $9_{0,1}$ | $7_{0,1}$ | | 1.3946 | 3.5832 | 3.2829 | 3.7794 | | 1.4073 | 3.4636 | 2.9585 | 3.8278 | 2.308 |
| $2_{0,2}$ | $2_{0,0}$ | 1.909 | 1.5677 | 1.5642 | 1.5710 | 1.5967 | 1.805 | 1.5728 | 1.5558 | 1.5511 | 1.5999 | 1.620 |
| $4_{0,2}$ | $4_{0,0}$ | | 0.3573 | 0.3507 | 0.3527 | 0.3698 | 0.406 | 0.3601 | 0.3466 | 0.3432 | 0.3715 | 0.348 |
| $6_{0,2}$ | $6_{0,0}$ | | 0.2110 | 0.2057 | 0.2158 | 0.2332 | | 0.2127 | 0.2027 | 0.2080 | 0.2348 | 0.198 |
| $3_{0,1}$ | $4_{0,0}$ | 0.664 | 1.2561 | 1.2497 | 1.2492 | 1.2976 | | 1.2657 | 1.2338 | 1.2102 | 1.3035 | 1.243 |
| $5_{0,1}$ | $6_{0,0}$ | | 1.1679 | 1.1410 | 1.0978 | 1.1843 | | 1.1848 | 1.1142 | 1.0317 | 1.1940 | 0.972 |
| $7_{0,1}$ | $8_{0,0}$ | | 1.3516 | 1.2257 | 1.0691 | 1.2038 | | 1.3793 | 1.1834 | 0.9696 | 1.2180 | 0.808 |
| $3_{0,1}$ | $2_{0,2}$ | 1.786 | 2.1539 | 2.1434 | 2.1488 | 2.2183 | | 2.1675 | 2.1212 | 2.0950 | 2.2267 | 2.171 |
| $5_{0,1}$ | $4_{0,2}$ | | 1.6359 | 1.5925 | 1.5145 | 1.6430 | | 1.6616 | 1.5519 | 1.4150 | 1.6576 | 1.313 |
| $7_{0,1}$ | $6_{0,2}$ | | 2.5884 | 2.1869 | 1.7760 | 2.0308 | | 2.6484 | 2.1005 | 1.5845 | 2.0575 | 1.260 |

Table 7
Same as Table 5 but for the $^{114}$Pd [81] and $^{116}$Pd [82].

| | | $^{114}$Pd | | | | | $^{116}$Pd | | | | | |
|---|---|---|---|---|---|---|---|---|---|---|---|---|
| $L^{(i)}_{n,n_w}$ | $L^{(f)}_{n,n_w}$ | Exp. | Z(5)-CFM | Z(5)-CFTH | Z(5)-CFE | Z(5)-CFK | Exp. | Z(5)-CFM | Z(5)-CFTH | Z(5)-CFE | Z(5)-CFK | Z(5) |
| $4_{0,0}$ | $2_{0,0}$ | | 1.5992 | 1.5196 | 1.4996 | 1.5052 | | 1.6030 | 1.5496 | 1.5450 | 1.6025 | 1.590 |
| $6_{0,0}$ | $4_{0,0}$ | | 2.3340 | 2.1172 | 2.0859 | 2.1055 | | 2.4627 | 2.2683 | 2.2417 | 2.4455 | 2.203 |
| $8_{0,0}$ | $6_{0,0}$ | | 2.9771 | 2.6108 | 2.6443 | 2.6897 | | 3.5895 | 3.0893 | 2.9915 | 3.4681 | 2.635 |
| $4_{0,2}$ | $2_{0,2}$ | | 0.7648 | 0.6872 | 0.6906 | 0.6939 | | 0.7626 | 0.6925 | 0.7015 | 0.7685 | 0.736 |
| $6_{0,2}$ | $4_{0,2}$ | | 1.1859 | 1.0237 | 1.0511 | 1.0687 | | 1.4142 | 1.1780 | 1.1685 | 1.3968 | 1.031 |
| $8_{0,2}$ | $6_{0,2}$ | | 1.9511 | 1.6525 | 1.8730 | 1.9312 | | 1.0439 | 2.3446 | 2.2638 | 2.9537 | 1.590 |
| $5_{0,1}$ | $3_{0,1}$ | | 1.3249 | 1.1857 | 1.1773 | 1.1888 | | 1.4061 | 1.2688 | 1.2597 | 1.3980 | 1.235 |
| $7_{0,1}$ | $5_{0,1}$ | | 2.1388 | 1.8527 | 1.9139 | 1.9514 | | 2.7250 | 2.2612 | 2.1867 | 2.6094 | 1.851 |
| $9_{0,1}$ | $7_{0,1}$ | | 2.8195 | 2.3970 | 2.7008 | 2.7864 | | 4.0390 | 3.5672 | 3.2975 | 4.2479 | 2.308 |
| $2_{0,2}$ | $2_{0,0}$ | | 1.6205 | 1.5596 | 1.5245 | 1.5300 | | 1.6228 | 1.5858 | 1.5773 | 1.6200 | 1.620 |
| $4_{0,2}$ | $4_{0,0}$ | | 0.3695 | 0.3322 | 0.3297 | 0.3326 | | 0.3858 | 0.3506 | 0.3493 | 0.3847 | 0.348 |
| $6_{0,2}$ | $6_{0,0}$ | | 0.2241 | 0.1930 | 0.1959 | 0.1983 | | 0.2423 | 0.2037 | 0.2088 | 0.2485 | 0.198 |
| $3_{0,1}$ | $4_{0,0}$ | | 1.2954 | 1.2052 | 1.1647 | 1.1750 | | 1.3536 | 1.2871 | 1.2641 | 1.3424 | 1.243 |
| $5_{0,1}$ | $6_{0,0}$ | | 1.0870 | 0.9689 | 0.9669 | 0.9843 | | 1.3160 | 1.1730 | 1.1184 | 1.2647 | 0.972 |
| $7_{0,1}$ | $8_{0,0}$ | | 0.9587 | 0.8307 | 0.8907 | 0.9165 | | 1.5508 | 1.2223 | 1.0968 | 1.3309 | 0.808 |
| $3_{0,1}$ | $2_{0,2}$ | | 2.2337 | 2.0914 | 2.0288 | 2.0433 | | 2.2939 | 2.1931 | 2.1656 | 2.2819 | 2.171 |
| $5_{0,1}$ | $4_{0,2}$ | | 1.4802 | 1.3157 | 1.3212 | 1.3472 | | 1.8531 | 1.6381 | 1.5483 | 1.7632 | 1.313 |
| $7_{0,1}$ | $6_{0,2}$ | | 1.5172 | 1.3093 | 1.4477 | 1.4969 | | 2.9186 | 2.1426 | 1.8386 | 2.2735 | 1.260 |

## 8. Conclusions

In this work, we used the CFNU method to investigate new analytical solutions of the CFBH with Morse, Tietz-Hua, and multi-parameter exponential potentials for triaxial nuclei and called the Z(5)-CFM, Z(5)-CFTH, and Z(5)-CFE models. The best values of the parameters of the potentials are determined by using the least-squares fitting process of the experimental energy spectra of the $^{114,116}$Pd, $^{126,128}$Xe, and $^{192,194}$Pt isotopes. Then the $B(E2)$ transition rates are determined using the same set of parameters. The numerical results demonstrate a significant degree of agreement with experimental findings. Between the three potentials that are studied in this work, the multi-parameter exponential-type potentials can be considered the best potentials to describe the triaxial nuclei. Using CFBH, the results of exponential-type potentials are identical to the results of Kratzer potential. The spectra of $^{192,194}$Pt isotopes, which are considered standard Z(5) CPS examples, are generated via the Z(5)-CFE model. The crucial point is that the CFBH can build multiple models that are similar to the original one, especially if $\alpha$ is extremely close to 1. Clearly, this idea is appealing since it allows the original models to be expanded in a variety of ways while still accurately fitting the experimental data without having to create a new model.